\documentclass[sigconf]{acmart}
\AtBeginDocument{%
  }

\copyrightyear{2026}
\acmYear{2026}
\setcopyright{cc}
\setcctype{by}
\acmConference[UIST '26]{The 39th Annual ACM Symposium on User Interface Software and Technology}{November 02--05, 2026}{Detroit, MI, USA}
\acmBooktitle{The 39th Annual ACM Symposium on User Interface Software and Technology (UIST '26), November 02--05, 2026, Detroit, MI, USA}
\acmDOI{10.1145/3830398.3830720}
\acmISBN{979-8-4007-2856-3/2026/11}

\usepackage{enumitem}

\newcommand{\pquote}[1]{\textit{``#1''}}

\usepackage{colortbl}   

\usepackage{algorithm}
\usepackage[noend]{algorithmic}

\usepackage{multirow}

\begin{document}

\title{WhatIf: Interactive Exploration of Large-Scale LLM Agent Simulations for Policy Reasoning}

\author{Yuxuan Li}
\affiliation{%
  \institution{School of Computer Science, Carnegie Mellon University}
  \city{Pittsburgh}
  \country{United States}
}
\email{yuxuanll@andrew.cmu.edu}

\author{Kyzyl Monteiro}
\affiliation{%
  \institution{School of Computer Science, Carnegie Mellon University}
  \city{Pittsburgh}
  \country{United States}
}
\email{kyzyl@cmu.edu}

\author{Hirokazu Shirado}
\affiliation{%
  \institution{School of Computer Science, Carnegie Mellon University}
  \city{Pittsburgh}
  \country{United States}
}
\email{shirado@cmu.edu}

\author{Sauvik Das}
\affiliation{%
  \institution{School of Computer Science, Carnegie Mellon University}
  \city{Pittsburgh}
  \country{United States}
}
\email{sauvik@cmu.edu}

\renewcommand{\shortauthors}{Li et al.}

\begin{abstract}

Policymakers in domains such as emergency management, public health, and urban planning must make decisions under deep uncertainty, where outcomes depend on how large populations interpret information, coordinate, and adopt over time.
Existing tools only partially support this process: tabletop exercises enable collaborative discussion but lack dynamic feedback, while computational simulations capture population dynamics but are designed for offline analysis.
We present \textsc{WhatIf}, an interactive system that enables policymakers to steer, inspect, and compare LLM-powered social simulations in real time.
Informed by a formative study in emergency preparedness planning, we derive four design requirements for interactive policy simulations: fluid steering, real-time scale, collaborative exploration, and multi-level interpretability.
We developed \textsc{WhatIf} guided by these requirements and evaluated it with five preparedness professionals across three disaster evacuation scenarios.
Our findings show that participants used the system as a space for iterative branching and comparison rather than evaluating fixed plans; reflected on tacit planning assumptions when agent behavior violated expectations; surfaced previously unrecognized planning vulnerabilities; and grounded their reasoning in inspectable agent-level cases rather than aggregate outputs alone.
These findings suggest broader design implications for LLM-powered social simulation systems: designing such systems as interactive, shared reasoning environments—rather than offline predictive tools—can better support expert decision-making under deep uncertainty.
\end{abstract}


\begin{teaserfigure}
  \includegraphics[width=0.943\textwidth]{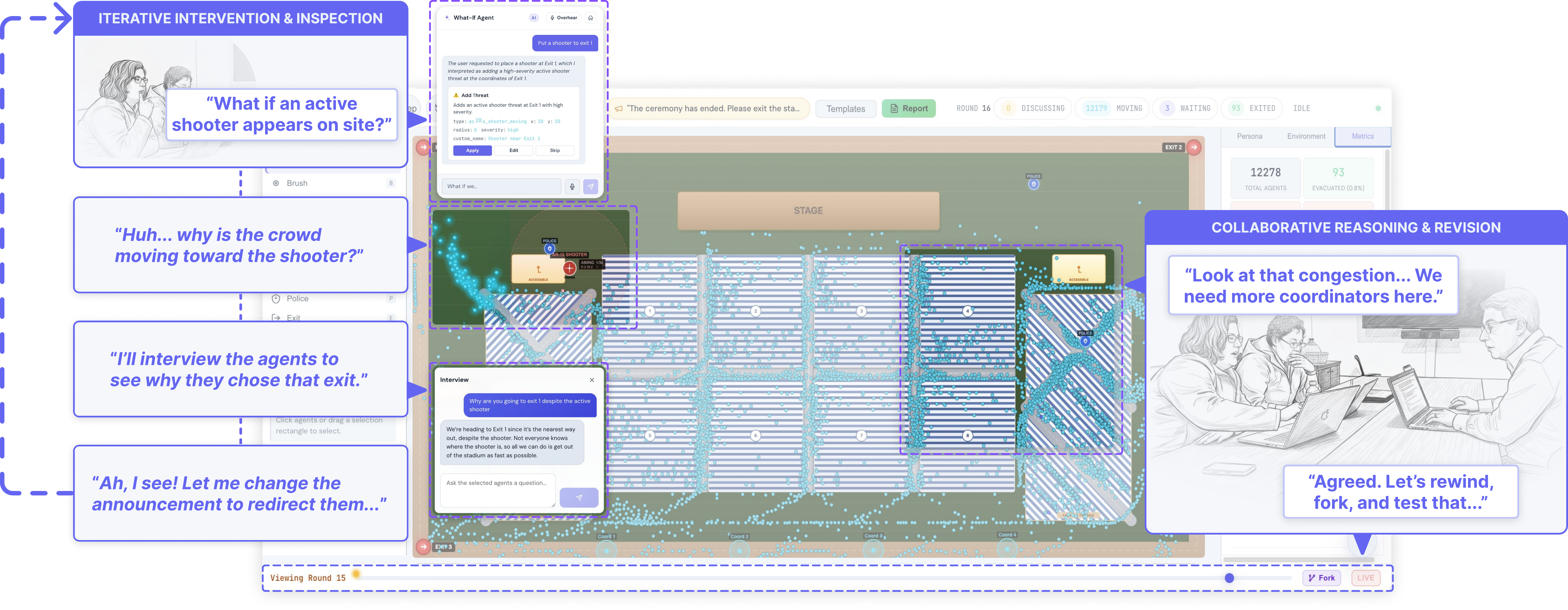}
  \caption{WhatIf enables policymakers to interactively explore large-scale LLM-powered social simulations as shared reasoning environments.
  Users can intervene in unfolding scenarios, inspect emergent behaviors at both agent and crowd levels, and compare branching outcomes in organizational policy planning.
    By linking individual decision rationales to system-level dynamics, the system supports real-time, what-if exploration and reasoning under uncertainty.}
  \Description{The figure illustrates the WhatIf system for interactive, collaborative exploration of large-scale LLM agent simulations. On the left, a sequence shows users posing a what-if question (e.g., introducing an active shooter), observing unexpected crowd behavior, and inspecting agent decisions through interviews. The center shows a large-scale simulation with thousands of agents moving through a venue, including highlighted regions of interest and an inspection panel displaying agent reasoning. On the right, multiple users collaboratively discuss and revise the scenario, proposing interventions (e.g., placing coordinators) and iterating on the simulation. The overall workflow depicts an iterative loop of intervention, inspection, and comparison across branching scenarios, supported by real-time interaction and shared workspace features.}
  \label{fig:teaser}
\end{teaserfigure}


\maketitle

\section{Introduction}

What if an active shooter appears at the school?
What if students misunderstand our warning message and panic?
What if we hired more safety coordinators instead of armed guards?

These are the kinds of questions policymakers and domain experts consider when making tactical policy decisions under deep uncertainty: i.e., situations, like emergencies, where outcomes, causal relationships, and even the relevant variables cannot be reliably specified in advance~\cite{walker2012deep, marchau2019decision, lempert2002agent}.
Across domains such as public health, urban planning, transportation, and emergency management, experts must anticipate how prospective interventions may play out as large populations interpret information, coordinate with one another, and adapt as events unfold.
For decision support systems, the challenge is not simply to model such situations, but to make them \textit{explorable}: helping policymakers compare alternatives, probe consequences, and reason through uncertainty with others.

Fundamentally, this challenge is as much an interaction problem as a modeling one. 
In such settings, decision-making is rarely about identifying a single optimal plan.
Instead, it involves collaboratively exploring uncertain possibilities, probing consequences, comparing alternatives, and revising assumptions as new considerations emerge.
Existing tools support only parts of this reasoning process. 
Tabletop exercises are flexible and collaborative, but lack dynamic feedback at realistic population scale~\cite{dausey2007designing}.
Computational simulations, including agent-based models, capture population dynamics, but are typically designed for offline analysis and post-hoc inspection through aggregate outputs~\cite{pelzer2014addedvalue,nespeca2023abm}.

Recent advances in LLM-powered agents offer a new opportunity to address this gap. 
Unlike conventional simulation agents, LLM agents can interpret ambiguous information, communicate with others, and produce inspectable rationales for their behavior. 
These capabilities make it possible to simulate not only where people move, but how they \textit{interpret, deliberate, and coordinate}—capabilities that are central to policy reasoning. 
At the same time, these more open-ended simulations are harder to validate and interpret, making interaction and inspection even more essential.

To explore this design space, we present \textsc{WhatIf}, an interactive system that enables policymakers to steer, inspect, and compare LLM-powered social simulations in real time while tracking how individual agent behaviors contribute to system-level outcomes.
We ground this work in disaster evacuation planning, a high-stakes domain in which officials must reason about how large, diverse crowds respond to changing threats, instructions, and infrastructure constraints.

We developed \textsc{WhatIf} through iterative engagement with emergency preparedness professionals and simulation prototypes, from which we derived four design requirements for interactive policy simulation systems: \textit{fluid steering}, \textit{real-time scale}, \textit{collaborative exploration}, and \textit{multi-level interpretability}.
Guided by these requirements, \textsc{WhatIf} enables policymakers to steer simulations through direct manipulation and natural language; explore over 12,000 LLM-powered agents at interactive speeds; collaborate synchronously in a shared workspace; and inspect outcomes ranging from agent rationales about individual decisions (e.g., ``Why did you run in this direction?'') to aggregate cross-run comparisons (e.g., casualty rates).

We evaluated \textsc{WhatIf} with five domain experts across disaster evacuation scenarios reflecting their primary planning concerns.
Our findings show that interactive LLM-powered social simulations support a distinct mode of policy reasoning centered on iterative branching, inspection, and collaborative deliberation rather than one-off plan evaluation.
Participants used the system to compare alternatives, interrogate unexpected behaviors, surface previously unrecognized planning vulnerabilities, and build arguments from inspectable agent-level cases.
Together, these findings suggest that the key opportunity of LLM-powered social simulation lies in enabling a long-missing interaction paradigm for expert reasoning under deep uncertainty.

This paper makes three contributions:
\begin{enumerate}[leftmargin=*,topsep=4pt,itemsep=2pt]
\item \textbf{Design requirements} for interactive what-if exploration of large-scale social simulations, derived from formative engagement with emergency preparedness professionals;
\item \textbf{\textsc{WhatIf}}, an interactive system that operationalizes these requirements for real-time steering, inspection, and comparison of LLM agent simulations at scale;
\item \textbf{Insights} into how domain experts use interactive simulations as shared reasoning environments for policy exploration, and the implications of these findings for the broader design of simulation-based decision support tools.
\end{enumerate}
\section{Related Work}

\subsection{Supporting Policy Under Deep Uncertainty}

Policy planning under deep uncertainty has motivated a long line of work on decision support and scenario exploration tools \cite{lempert2002agent}.
\textit{Deep uncertainty} refers to decision contexts in which threats, outcomes, causal relationships, and even the relevant variables cannot be reliably specified in advance \cite{walker2012deep, marchau2019decision}.
Under such conditions, support systems aim to help practitioners explore scenarios and visualize consequences, though adoption has often been limited by mismatches between system capabilities and how experts actually reason in practice~\cite{geertman2008pss,pelzer2014addedvalue,klosterman1999if}.
Related work in what-if analysis and visual analytics has similarly emphasized interactive manipulation of model inputs, comparison of alternatives, and iterative hypothesis refinement~\cite{wexler2019if,reinert2020visual,DasAntar2024VIMEVI,endert2014human}.
These systems establish the importance of interaction for reasoning with uncertainty, but they are not typically designed to support collaborative exploration of large-scale social dynamics as they unfold.

In crisis management specifically, tabletop exercises remain a primary tool for collaborative rehearsal, coordination, and role clarification~\cite{dausey2007designing,solinska2018overview,radianti2018co}.
Participatory simulations extend this tradition by embedding people within modeled systems, enabling experiential understanding of emergent phenomena~\cite{colella1998participatory,wilensky1999participatory}.
These approaches support discussion and shared sensemaking, but provide limited support for dynamically exploring how interventions propagate through large populations over time. 
 
Computational social simulations, such as agent-based models, provide a complementary computational foundation \cite{lempert2002agent}.
Prior work has used simulations to study evacuation behavior, crowd dynamics, and other complex social systems in which system-level outcomes emerge from local interactions~\cite{macal2005tutorial,senanayake2024agent,helbing1995social,perez2009agent}.
However, most such systems are built for offline analysis rather than interactive use, offering limited support for steering simulations during execution, inspecting individual agent behavior or decision rationale, or collaborating around a shared evolving simulation ~\cite{senanayake2024agent,mulder1998computational,geertman2020planning}.
 
Taken together, these strands of work highlight three important but rarely integrated capabilities for policy reasoning under uncertainty: interactive exploration, collaborative sensemaking, and dynamic simulation of population-level behavior.
\textsc{WhatIf} brings these capabilities together by treating computational social simulation not as an offline predictive tool, but as a shared, interactive reasoning environment that allows for synchronous, collaborative policy reasoning.

\subsection{LLM-Powered Social Simulations}

Recent advances in large language models have opened new possibilities for simulating human behavior.
Compared with rule-based agents, LLM-powered agents can more flexibly represent open-ended decision-making, communication, and social interaction, and have been used to construct generative agents, role-based simulations, and synthetic social environments~\cite{park2023generative,li2023camel,gao2024large,Xie2024AIMS}.
This expressiveness is especially promising for domains in which behavior depends on ambiguous instructions, social relationships, and contextual interpretation rather than fixed response rules~\cite{dang2025large,lindell2012protective,kinateder2015risk}.

However, most prior work on LLM-powered social simulation has focused on agent architectures, prompting strategies, and retrospective analysis~\cite{li2023camel,Liu2023AgentBenchEL,Xie2024AIMS,NEURIPS2024_1cb57fcf}.
As a result, relatively little attention has been paid to the interaction layer required for expert use: how domain experts can steer simulations while they run, inspect why agents behave as they do, and use those observations to guide further exploration and policy reasoning~\cite{gao2024large,gurcan2024llm,li2025makes,Zeng2026TooHT}.

Prior HCI work on interpretable AI~\cite{amershi2019guidelines,ehsan2023charting}, human-in-the-loop simulation~\cite{mulder1998computational,bavoil2005vistrails,cutler2020trrack}, and collaborative visual analytics~\cite{viegas2007manyeyes,isenberg2011co,heer2007voyagers} highlights the importance of runtime steering, multi-level explanation, and shared sensemaking.
Our work brings these concerns into the context of LLM-powered social simulation by focusing on the interaction and system design challenges that arise when such simulations are used as real-time reasoning environments for policy exploration. 
\section{Formative Study} \label{sec:formative}

To inform the design of interactive systems for what-if exploration under uncertainty, we conducted a formative study with emergency preparedness professionals responsible for emergency preparedness and planning at a medium-sized university.
From this study, we derived four design requirements for making social simulations useful in policy practice.
 
\subsection{Study Context and Approach}
We worked with a team of five emergency preparedness professionals responsible for evacuation protocol design, public communication, incident monitoring, and cross-team coordination at a university.
This collaboration allowed us to examine policy decision-making as it unfolds through shared organizational responsibilities and coordination practices \cite{maitlis2005social}.
To anchor feedback in a concrete use case, we focused on emergency evacuation planning for a large outdoor commencement ceremony with approximately 12,000 attendees---a setting participants described as operationally consequential because it combines heterogeneous crowds, coordination across roles, and rapidly changing conditions.
 
The study spanned six design iterations and three semi-structured interviews (1.5--2 hours each) over 16 months.
We used early simulation prototypes as technology probes~\cite{hutchinson2003technology} to elicit reactions, surface latent needs, and ground discussion in concrete planning scenarios.
Two researchers then conducted a thematic analysis~\cite{Braun01012006} of the interview transcripts, iteratively coding for current planning practices, forms of uncertainty, breakdowns in existing methods, desired interaction capabilities, and interpretability needs.
All study procedures were approved by the university's Institutional Review Board (IRB).

\subsection{Design Requirements}
 
Our analysis surfaced four recurring needs (see Appendix Section~\ref{app:formative} for representative quotes), which we translate into design requirements for interactive social simulations in policy use.  
These requirements emerged from triangulating across sources and interview sessions: participants' explicitly stated needs, limitations we observed in their current planning practice, and their reactions to the simulation probes together pointed to the same recurring capabilities.

\paragraph{DR1: Fluid steering.}
Intervention planning is fundamentally exploratory and shaped by uncertainty in how people interpret information and situations.
The team described their work as exploring multiple possible futures: e.g., probing how different groups might interpret alerts, how people might react to visible threats, and how local decisions could cascade into crowd-level effects.
During prototype walkthroughs, team members immediately proposed testing alternative wordings, suggesting a need for fast, flexible intervention where users can modify conditions and immediately observe consequences.
This motivated \textbf{(DR1) fluid steering:} systems should support rapid, low-friction modification of simulation conditions with immediate feedback.

\paragraph{DR2: Real-time scale.}
The team needed dynamic exploration at population scale, but existing planning methods did not support it.
They described their current planning approach as relying on tabletop exercises, meetings, and templates.
These practices help build shared mental models and clarify roles, but provide limited support for observing how behavior unfolds dynamically at crowd scale.
Several team members also emphasized that simulation must operate quickly enough to support live collaborative reasoning---if results take hours, the planning group has already disbanded.
These findings motivated \textbf{(DR2) real-time scale:} systems should remain interactive while simulating thousands of agents, so that emergent dynamics become visible during live exploration.

\paragraph{DR3: Collaborative exploration.}
Policy planning is inherently collaborative, and simulation should support shared exploration.
The team consistently described planning as distributed across roles.
Different stakeholders contribute different expertise: some monitor public discourse, some manage protocols, others maintain situational awareness.
These findings motivated \textbf{(DR3) collaborative exploration:} simulations should be considered shared artifacts around which stakeholders can, in real-time, jointly inspect assumptions, intervene, and discuss consequences.
 
\paragraph{DR4: Multi-level interpretability.}
System-level outputs are insufficient without inspectable explanations of agent behavior.
Although the team cared about system-level outcomes such as congestion patterns and evacuation rates, they did not treat such outputs as sufficient for decision-making.
What repeatedly drove discussion was the need to understand \emph{why} a particular behavior occurred.
When agents misread announcements or chose unexpected routes, some team members wanted to inspect why and use that insight to generate new hypotheses about message phrasing or resource placement.
These findings motivated \textbf{multi-level interpretability (DR4):} systems should facilitate inspection and sensemaking at multiple scales --- from individual agent reasoning to population-level outcomes.

\begin{figure*}[ht]
    \centering
    \includegraphics[width=0.90\linewidth]{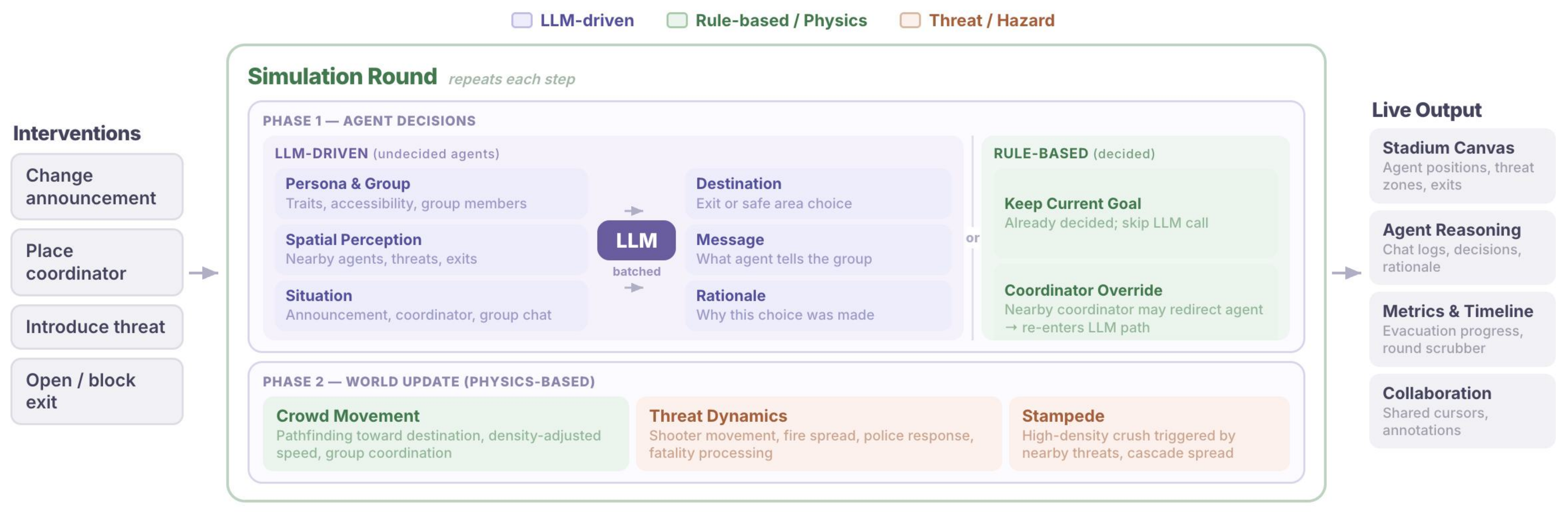}
    \caption{\textsc{WhatIf} system overview. After user-authored interventions (left), the system executes repeated simulation rounds that combines LLM-driven agent decisions with physics-based world updates (center). Interventions update shared state and agent context. 
    Agents enter LLM deliberation when their context changes, producing destinations, messages, and rationales, while a deterministic engine advances movement, hazards, and density effects. Outputs are reflected in synchronized views (right), enabling users to inspect how interventions propagate from individual decisions to crowd-level outcomes in real time.
    }
    \Description{A diagram illustrating the simulation loop. On the left, user interventions (editing announcements, placing coordinators, introducing threats, opening or blocking exits) feed into a repeated simulation round. The round has two phases: Phase 1 (Agent Decisions), where agents either follow rule-based shortcuts or use an LLM to process persona, spatial perception, and situational inputs to produce a destination, message, and rationale; and Phase 2 (World Update), where a physics-based engine updates crowd movement, threat dynamics, and density effects such as stampedes. On the right, outputs are shown in multiple synchronized views, including a spatial canvas of agents and threats, agent reasoning traces, metrics and timeline, and collaborative annotations.}
    \label{fig:dataflow}
\end{figure*}

\section{The WhatIf System}
 
Guided by these four design requirements, \textsc{WhatIf} is an interactive system for exploring, steering, and comparing large-scale LLM-powered social simulations in real time.
Rather than treating simulation as an offline predictive tool, \textsc{WhatIf} is designed as a shared reasoning environment in which policymakers can intervene in unfolding scenarios, inspect agent behavior, and collaboratively explore alternatives. 
Although we developed and evaluated the system in the context of disaster evacuation planning for a university commencement, its architecture is designed to generalize across different populations, spatial settings, and threat scenarios; our empirical findings, however, come from this single context.

\subsection{System Overview}

\textsc{WhatIf} combines two computational layers: an \emph{LLM decision layer} that drives agent deliberation, and a \emph{deterministic spatial engine} that handles movement, hazard evolution, and crowd dynamics.
The LLM determines \emph{where} an agent wants to go and \emph{why}; the spatial engine determines \emph{how} that agent physically moves.
This separation keeps agent behavior expressive and inspectable while maintaining interactive performance.
 
The system simulates a population of socially situated agents, each with a persona, group membership, and evolving local context.
When new information enters an agent's context---such as a revised announcement, a nearby threat, or a coordinator's directive---the agent enters a deliberation phase in which the LLM produces a destination choice, a natural-language rationale, and, when appropriate, a message to its social group.
Each agent is endowed with a persona and group membership and reasons over a bounded local context---its recent interaction history, nearby people and threats, and a ranked set of reachable exits---rather than maintaining a long-term memory stream as in prior generative-agent architectures~\cite{park2023generative}.
This is a deliberate simplification: restricting each agent's context, and invoking the LLM (\texttt{gpt-4.1-nano}) only when that context changes, is what keeps the system interactive at a population scale of over 12{,}000 agents.
We describe agent context construction, batched deliberation, and history handling in detail in Appendix~\ref{app:implementation}.

Users modify unfolding scenarios through \emph{interventions}, including changes to announcements, personnel placement, exits, obstacles, and threats.
All interventions are applied through a unified engine that keeps simulation state synchronized across users and input modalities.
 
Simulation proceeds in repeated rounds.
In each round, agents whose context has changed enter batched LLM deliberation, while agents with stable context continue under the deterministic engine.
The spatial engine then advances movement, hazard dynamics, and density effects, with outputs continuously reflected in the interface (Figure~\ref{fig:dataflow}).

For the evaluation reported in this paper, we instantiated \textsc{WhatIf} with a population of 12,278 agents representing attendees at a large outdoor commencement ceremony.
This instantiation was derived from formative engagement with the planning team, but the system is not specific to this population or scenario.
Instantiating a new setting requires generating a persona-conditioned agent population, supplying a spatial map and navigation grid, and parameterizing the relevant threats (Appendix~\ref{app:design-choices}); we have not yet empirically validated the system beyond this single context.

\subsection{Design Features}
\subsubsection{\textbf{Fluid Steering (DR1)}}

\begin{figure*}[ht]
    \centering
    \includegraphics[width=0.95\linewidth]{figures/steering.png}
    \caption{Fluid steering in \textsc{WhatIf}. Users modify unfolding scenarios by selecting a placement tool and clicking directly on the canvas; each callout pairs a tool's configuration panel with its resulting effect on the canvas: coordinators, with a target exit and optional custom message broadcast to nearby agents (1); threats spanning severe weather, fire, chemical, shooter, and explosive hazards, with the selected threat's modeled behavior exposed to the user (2); police (3); custom exits (4); and obstacles (5). Announcements can be edited in place or drawn from reusable templates (6), and the environment panel exposes all active scenario parameters for direct editing and removal (7). The What-If Agent (8) translates typed or spoken requests into structured interventions that users preview and then apply, edit, or skip. All interventions flow through a shared engine, so changes are immediately reflected in agent behavior and crowd dynamics.}
    \Description{An annotated screenshot of the WhatIf interface with eight numbered callouts. The central canvas shows a stadium venue with thousands of cyan agent dots moving through seating blocks toward exits. Callouts 1--5 each pair a configuration panel with the corresponding element placed on the canvas: a coordinator panel with target-exit dropdown and custom message field, alongside four coordinators positioned along the bottom of the canvas (1); a threat palette of six hazard types and a detail panel describing the selected active shooter's weapon and movement behavior, alongside a magnified view of the shooter and nearby police (2); the police, exit, and obstacle tools in the placement toolbar, alongside a police unit, a custom garage exit, and a fallen-tree obstacle (3--5). Callout 6 marks an editable announcement field and templates button in the top bar. Callout 7 marks a right-hand environment panel listing the announcement, coordinators with assigned exits, the active threat, police with coordinates, and custom exits, each with edit and delete controls. Callout 8 marks the What-If Agent panel, where a typed request to place coordinators near the stage has produced two structured commands with concrete coordinates and messages, each with apply, edit, and skip buttons.}
    \label{fig:steering}
\end{figure*}

To support fluid steering, the system enables rapid, low-friction modification of simulation conditions with immediate feedback.
\textsc{WhatIf} supports fluid steering through a unified intervention model that allows users to modify unfolding scenarios and immediately observe their effects (Figure~\ref{fig:steering}).

\textit{Direct manipulation.}
On the central canvas, users directly modify simulation elements---such as coordinators, exits, obstacles, and hazard sources (grounded in FEMA's all-hazards framework~\cite{goss1998guide})---through simple interactions such as clicking and dragging.
They can also reposition individual agents or groups and edit public announcements in place. 
When an intervention changes agents’ context (e.g., updating an announcement), affected agents re-enter deliberation, allowing users to observe how new instructions propagate through a heterogeneous population.

\textit{Natural-language input.}
To reduce steering effort, users can also issue commands using natural language.
A ``What-If Agent'' translates these inputs (text/voice) into structured interventions, enabling users to express scenario changes in natural language while remaining consistent with direct manipulation.
This approach builds on prior work showing natural language queries can map to structured analytic specifications without requiring knowledge of the underlying representation \cite{Narechania2020NL4DVAT}.

All forms of steering flow through a shared intervention engine, ensuring that changes are synchronized across users, reversible through the timeline, and immediately reflected in the simulation.

\subsubsection{\textbf{Real-Time Scale (DR2)}}

To support real-time scale, the system remains interactive while simulating thousands of agents, so that emergent dynamics become visible during live exploration.
Planning questions about bottleneck formation, exit imbalances, and cascading effects only become visible at realistic population scales.
\textsc{WhatIf} maintains interactive performance with over 12,000 agents by combining selective LLM use with efficient spatial simulation.

\textit{Selective deliberation.}
Inspired by prior work~\cite{Park2011LargeSC}, we design agents to invoke the LLM only when their context changes (e.g., new announcements or nearby threats).
Otherwise, they continue under the deterministic spatial engine, substantially reducing the number of LLM calls while preserving responsiveness.
We ground the spatial engine’s parameters in academic papers, reports, and incident data, as well as expert knowledge identified during the formative study.
See Appendix Section~\ref{app:design-choices} for the design choices and their empirical grounding.

\textit{Batched processing and spatial efficiency.}
LLM requests are batched across agents. 
Prior work has shown that this approach enables large-scale LLM agent simulations with minimal performance loss~\cite{Piao2025AgentSocietyLS, Yan2025AgentSocietyCD}.
Movement is handled by a precomputed flow-based spatial model with extensive caching.

Together, these allow full simulation rounds to complete within seconds in typical cases, enabling users to iteratively explore scenarios without interrupting collaborative reasoning.

Under default settings (150 rounds), most rounds complete in approximately 1 second, with occasional slower rounds when many agents re-deliberate simultaneously (Figure~\ref{fig:round_time}).

\subsubsection{\textbf{Collaborative Exploration (DR3)}}

\begin{figure}[ht]
    \centering
    \includegraphics[width=\linewidth]{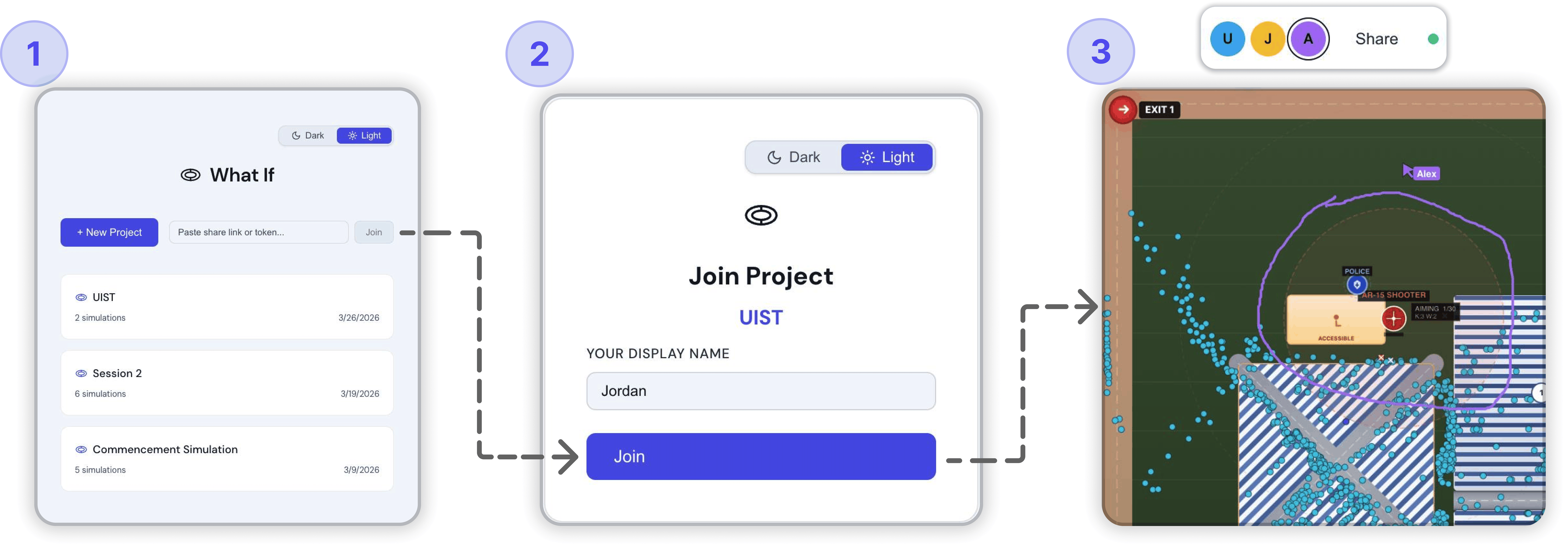}
    \caption{Collaborative exploration in \textsc{WhatIf}. Users access shared project workspaces with multiple simulation runs (1) and join live session through a lightweight ``Join Project'' flow (2). 
    Once connected, collaborators share a synchronized simulation state with visible presence through avatars and live cursors (3). A shared interactive canvas allows users to inspect, annotate, and intervene in the same unfolding simulation together (4), supporting real-time discussion and coordinated exploration.}
    \Description{A three-panel interface illustrating collaborative interaction in a simulation system. The left panel shows a project workspace listing multiple simulation scenarios. The middle panel shows a “Join Project” dialog where a user enters a display name to join a shared session. The right panel shows the main simulation canvas with multiple users present, indicated by avatars and cursors. The canvas displays a crowd simulation with annotations and drawn markings, demonstrating that all participants can view, interact with, and modify the same simulation state in real time.}
    \label{fig:collaboration}
\end{figure}

To support collaborative exploration, the system enables multiple stakeholders to explore and discuss the same simulation in real time.
Policy planning is often distributed across roles, and no single expert's perspective is sufficient.
\textsc{WhatIf} treats the simulation as a shared artifact in which stakeholders can jointly inspect, intervene, and discuss unfolding scenarios (Figure~\ref{fig:collaboration}).

\textit{Shared sessions.}
Users join a shared session in which all simulation state, interventions, and interaction events are synchronized.
Changes made by any user are immediately reflected for all collaborators, allowing participants to observe and reason about the same evolving scenario.

\textit{Live presence and annotation.}
The interface makes collaborators’ actions visible in real time through cursor presence and shared annotations.
Users can highlight areas of concern, sketch alternative plans, or point to emerging patterns directly on the simulation.
These interactions support the kinds of informal, situated reasoning found in tabletop exercises, now grounded in a dynamic simulation.

\textit{Synchronized state.}
All interventions apply to a single shared simulation instance.
When one user modifies the scenario—such as repositioning coordinators or editing an announcement—others immediately observe the resulting changes and their downstream effects.
This shared state enables coordinated exploration and discussion of alternative interventions.

\subsubsection{\textbf{Multi-Level Interpretability (DR4)}}

\begin{figure*}[ht]
    \centering
    \includegraphics[width=0.95\linewidth]{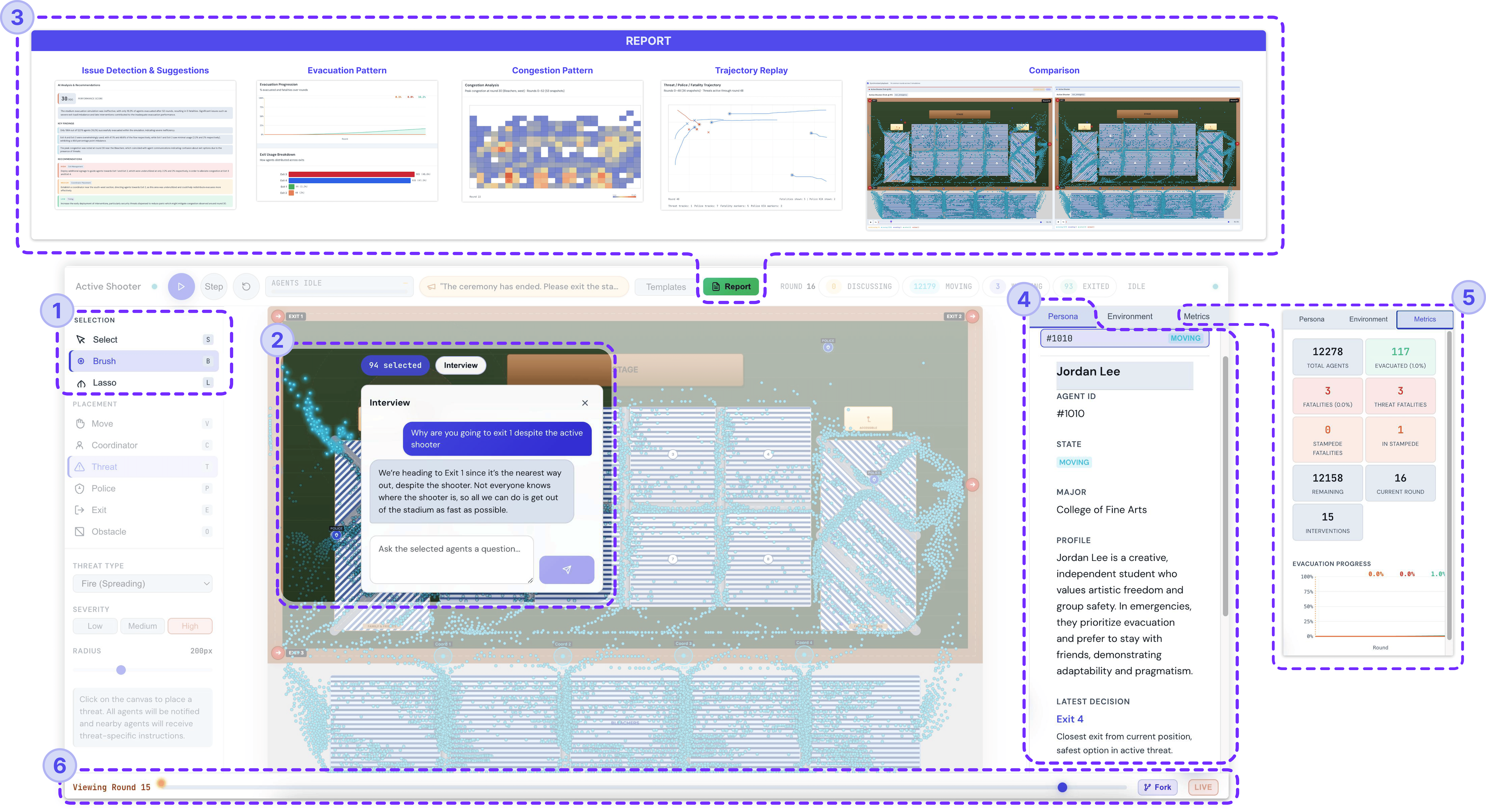}
    \caption{Multi-level interpretability in \textsc{WhatIf}. \textsc{WhatIf} supports inspection across levels of abstraction and time, linking individual agent reasoning to system-level outcomes. (1) A report view summarizes outcomes---including issue detection, evacuation progress, congestion patterns, and trajectory comparisons---to support post-hoc analysis and cross-run evaluation. (2) In-character interviews expose agents’ situated reasoning behind decisions. (3) The live canvas visualizes large-scale crowd dynamics, allowing users to relate local actions to emergent system behavior. (4) An inspection panel reveals agent personas, states, rationales, and group communication. (5) System-level metrics provide real-time summaries of population outcomes during execution. (6) Temporal navigation and replay enable rewinding, branching, and comparing alternative scenarios for counterfactual analysis.}
    \Description{A screenshot of the WhatIf interface highlighting multi-level interpretability features. A central canvas shows thousands of simulated agents moving through a venue, with dense trajectories indicating crowd flow. A pop-up interview dialog overlays selected agents, showing a natural-language response explaining their decision (e.g., exit choice). On the right, an inspection panel displays an individual agent’s persona, current state, rationale, and profile details. Adjacent panels show system-level metrics such as total agents, evacuation counts, fatalities, and progress over time. At the top, a reporting dashboard includes visual summaries such as evacuation progress charts, congestion heatmaps, trajectory replays, and side-by-side comparisons of simulation runs. A timeline at the bottom enables stepping through rounds, replaying, and branching scenarios. The layout emphasizes connections between individual agent reasoning and aggregate system outcomes.}
    \label{fig:interpretability}
\end{figure*}

To support multi-level interpretability, the system enables inspection from individual agent reasoning to system-level outcomes.
Policymakers do not treat system-level outcomes as sufficient for decision-making.
\textsc{WhatIf} supports reasoning by allowing users to move across levels of abstraction and across time, connecting individual behavior to collective outcomes (Figure~\ref{fig:interpretability}).
 
\textit{From agents to system outcomes.}
Users can inspect individual agents to understand how local decisions contribute to system-level effects.
Clicking an agent reveals its persona, current state, and natural-language rationale from its most recent deliberation, along with group-level communication.
Users can also query individual agents or groups directly, receiving in-character responses grounded in their local context.
Prior work shows that interviewing agents and examining their rationales are effective methods for interpreting their behavior~\cite{Park2024GenerativeAS, li2025actions, Li2026HowWC, 10.1145/3746059.3747696}.
Together, these views make agent behavior inspectable rather than opaque.

\textit{System-level visualization.}
At the same time, the interface presents system-level outcomes such as evacuation progress, per-exit throughput, congestion patterns, and agent state distributions.
We selected these summaries because they correspond to the system-level outcomes the planning team identified as decision-relevant during the formative study (Section~\ref{sec:formative}), such as congestion patterns and evacuation rates.
The canvas integrates these signals, allowing users to relate individual decisions to emerging system-level dynamics.

\textit{Temporal navigation and counterfactual exploration.}
Users can navigate the simulation timeline, inspect past states, and branch from any prior moment to explore alternative interventions.
This supports direct counterfactual reasoning by comparing how different decisions lead to diverging outcomes from the same starting point.

\textit{Comparison across runs.}
To support systematic evaluation, users can compare multiple simulation runs side by side, examining differences in trajectories and outcomes across alternative interventions.

\subsection{Implementation}

\textsc{WhatIf} is implemented as a web application with a Python backend (FastAPI, async I/O, SQLite) and a React frontend.
Additional implementation details, including system architecture and performance characteristics, are provided in Appendix Section~\ref{app:implementation}.

Together, these features make \textsc{WhatIf} more than a simulation viewer or scenario editor. 
By combining real-time steering, scalable execution, shared exploration, and multi-level inspection, the system supports the use of large-scale social simulations as interactive reasoning environments in which policymakers can collaboratively probe uncertainty, test alternatives, and interpret how local behaviors give rise to system-level outcomes.

\section{Evaluation}

We evaluated \textsc{WhatIf} with five emergency preparedness professionals across three disaster evacuation scenarios.
These scenarios were selected to reflect issues and planning concerns that repeatedly surfaced during the formative study.
Our goal was to understand how domain experts use interactive LLM-powered social simulations as reasoning environments---and, more broadly, what interaction patterns and design tensions emerge when such simulations are used for collaborative policy exploration in high-stakes settings.
 
\subsection{Participants}

We recruited five emergency preparedness professionals (P1--P5) from the same team involved in the formative study, all of whom have direct responsibilities for evacuation planning, public communication, and incident response.
P1--P3 had participated in the formative study interviews, whereas P4 and P5 were new to the system.
Their experience ranged from 3 to 18 years; only P1 had prior experience using simulation software (Arena, a discrete-event tool).

Because all participants belonged to the same organizational team, the evaluation captures how an actual planning group engages with interactive simulation in a shared institutional context.
This strengthens ecological validity---participants brought genuine shared responsibilities, organizational knowledge, and collaborative norms---but limits generalizability across organizations and planning cultures, a point we return to in the discussion.

\subsection{Study Design}

The evaluation comprised three task types designed to surface how participants use the system for guided familiarization, open-ended individual exploration, and shared collaborative reasoning.
 
\textit{T1: Guided severe weather response (30 min, individual).}
A researcher walked each participant through a step-by-step tornado evacuation scenario covering all major system features.
All participants completed T1.

\textit{T2: Open-ended active shooter response (45 min, individual).}
Participants independently developed evacuation protocols for an active shooter scenario involving the full 12,278-agent population, with access to all features except those related to collaboration.
All participants completed T2.

\textit{T3: Collaborative active shooter response (40 min, group).}
P1--P3 first reviewed and compared their individual T2 results using the comparison view (10 min), then worked together on a shared simulation (30 min) with collaboration features such as synchronized state, live cursors, and shared annotations.
P4 and P5 did not participate in T3 because they are not directly involved in collaborative evacuation planning events.

After each session, participants completed a questionnaire that included NASA-TLX items~\cite{hart1988development} (7-point scale), a 7-point adaptation of the System Usability Scale~\cite{brooke1996sus}, custom Likert-scale items aligned with the four design requirements (excluding questions related to collaboration and future adoption for P4 and P5), and open-ended questions.
We also conducted semi-structured interviews with all five participants individually after their sessions.

\subsection{Analysis}

Two researchers conducted a thematic analysis~\cite{Braun01012006} of all session transcripts and interview recordings, iteratively coding participant utterances into categories spanning interaction behavior, sensemaking, collaboration, domain reasoning, and validity judgments, then consolidating codes into interpretive themes.

We complemented this analysis with descriptive quantitative summaries from questionnaires, interaction logs, and simulation records.
We exclude T1 from quantitative analysis because participants' actions were largely guided by the researcher.
Given the small sample, we report individual scores and summary statistics to contextualize qualitative findings rather than performing inferential tests.

\subsection{Findings}

Our analysis reveals five recurring interaction patterns in how domain experts used \textsc{WhatIf} as a shared reasoning environment: iterative branching, reflection through expectation violation, surfacing previously unseen planning vulnerabilities, reliance on inspectable cases over aggregate outputs, and shared deliberation around a common simulation artifact.
These findings point to broader design implications for systems that support simulation-based policy exploration.

\subsubsection{Participants Explored ``What If'' Scenarios via Iterative Branching}

\begin{figure*}[ht]
    \centering
    \includegraphics[width=0.75\linewidth]{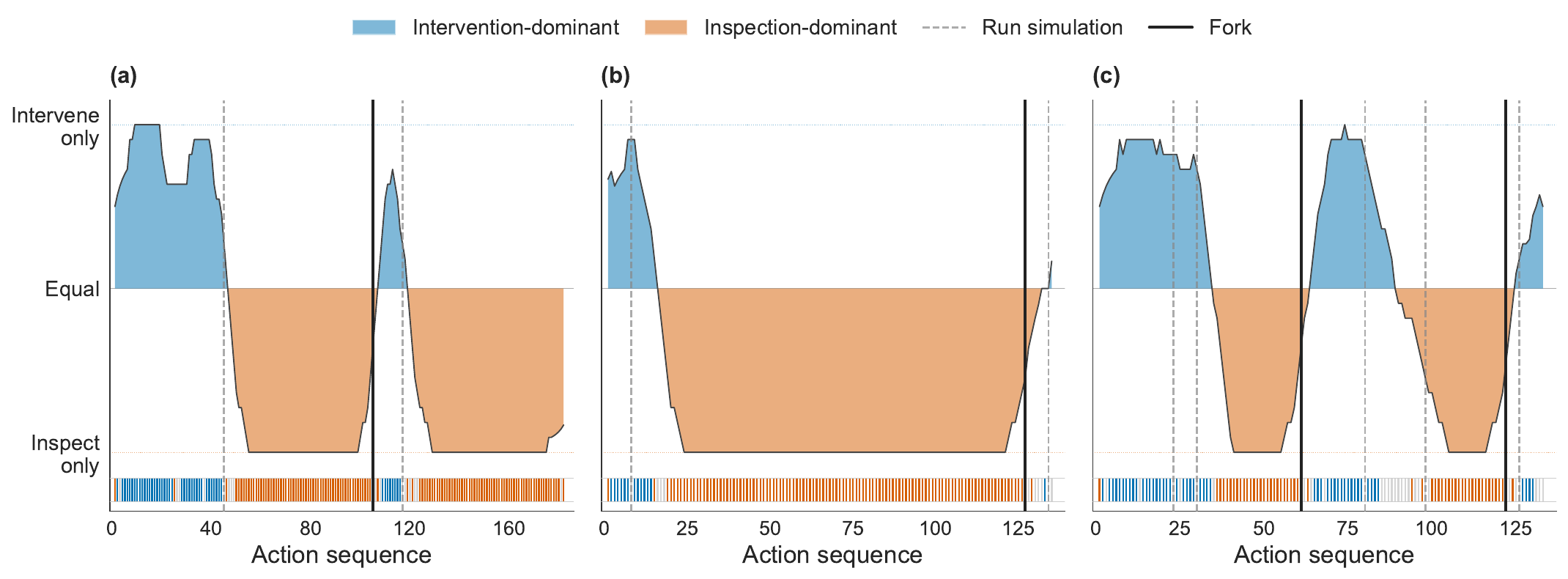}
    \caption{Iterative what-if cycles across fork episodes. Each panel corresponds to a different participant (P2, P4, P5) and shows the rolling balance between intervention actions (blue, above zero) and inspection actions (orange, below zero) over a parent–child fork chain; dashed gray lines indicate simulation runs and solid black lines indicate fork points, and the bottom rug marks individual actions. (a) P2 shows a full intervene $\rightarrow$ simulate $\rightarrow$ inspect $\rightarrow$ fork cycle. (b) P4 is dominated by a sustained inspection with repeated simulation runs before a late fork. (c) P5 shows two successive cycles branching from the same base run, with intervention resuming after each run. Across participants, a recurring intervene--simulate--inspect--fork pattern emerges during exploration.}
    \Description{Three line charts (panels a–c) show how participant actions evolve over time during simulation sessions with branching (forks). The x-axis represents action sequence; the y-axis represents a balance between intervention actions (positive, blue) and inspection actions (negative, orange). Each chart includes dashed vertical lines marking simulation runs and solid vertical lines marking fork points. A strip along the bottom encodes individual actions as colored ticks. Panel (a) shows an early period dominated by interventions, followed by simulation runs, then a sustained shift to inspection before a fork, illustrating a full intervene--simulate--inspect--fork loop. Panel (b) is dominated by inspection throughout, with multiple simulation runs and a fork occurring late. Panel (c) shows two repeated cycles: intervention-dominant activity transitions to inspection, followed by forks, after which intervention resumes on each new branch. Across all panels, the dominant pattern is cyclical: participants intervene to modify the simulation, run it to observe outcomes, inspect results, and then fork to explore revised strategies.}
    \label{fig:user_study}
\end{figure*}

Participants did not use \textsc{WhatIf} as a tool for evaluating fixed plans.
Instead, they treated it as a space for iterative branching: placing interventions, observing consequences, diagnosing unexpected outcomes, and revising strategies through repeated what-if cycles.
P2 captured this pattern: \pquote{We get multiple what-ifs---what if we do this? What if we had more police officers? We could play with that.}
Similarly, P5 ran three simulation instances, adding coordinators after reviewing her first run's report.

Interaction logs corroborate this exploratory orientation.
Across five individual sessions, participants created 14 simulations, authored 259 interventions spanning coordinator placements, police deployments, obstacle additions, exit modifications, and announcement edits, and ran simulations for 593 total rounds.
Participants adopted different strategy repertoires: P2 and P3 favored coordinator-heavy approaches, while P4 relied predominantly on police placements and obstacles.

Forking and temporal navigation were central to exploration.
Three fork episodes occurred across participants, each exhibiting a recurring cycle: a burst of interventions, followed by a shift to diagnostic activity (examining agents' rationales, reviewing the timeline, conducting interviews), and then a fork to try a revised strategy (Figure~\ref{fig:user_study}).
This intervene--simulate--inspect--fork loop suggests that interactive simulation systems should support scenario exploration not as a linear workflow, but as one that centers iterative refinement and branching.

\subsubsection{Unexpected Agent Behavior Surfaced Tacit Planning Assumptions}

One of the most analytically productive moments in the study occurred when agent behavior violated participants’ expectations.
For example, P1 articulated: \pquote{If you already know where they're gonna go, why do we need the simulation?}
He recognized that his own expectations constituted a planning bias.
When agents ignored his announcement and continued toward an exit near a fire, he described three distinct reactions: frustration, self-awareness, and ultimately acceptance that the simulation had revealed something he had not considered.
This reflective quality was not explicitly designed into the system; it emerged from the combination of inspectable agent behavior and real-time feedback that made expectation violations immediately visible.

For interactive simulation systems more broadly, this finding suggests that \emph{inspectability of agent behavioral traces and rationales matters not only for understanding the simulation, but for surfacing the tacit knowledge experts bring to it.}
Unexpected behavior is to be expected when simulating scenarios under deep uncertainty. When the reasoning behind unexpected behavior is visible, however, experts were better positioned to reason about whether an unexpected behavior reflected simulation implausibility or a genuine planning blind spot.

\subsubsection{Previously Invisible Planning Vulnerabilities Revealed Through Interactive Exploration}
 
A consequential use of \textsc{WhatIf} was surfacing vulnerabilities the team had not previously recognized.
Most notably, participants realized that people with accessibility needs would be forced to cross the entire stadium under certain evacuation conditions because only one accessible exit was available.
P3 noticed agents with accessibility needs traveling across the entire stadium to reach exits equipped with ramps.
She found this unacceptable: \pquote{When there's a shooter, we can't have people crossing the entire field just because that's where the ramp is.}
This observation shifted the conversation from evacuation routing to infrastructure policy---a question the team had not raised in prior tabletop exercises.

The active shooter scenario also opened new reasoning spaces.
After watching a shooter breach the stadium, P1 began reasoning about differentiating responders by capability and positioning them strategically, moving planning from quantity of responders to composition and placement.
The same scenario prompted debate about whether to send any announcement at all during an active shooting, a question about timing and modality that would not have surfaced without the simulation making the constraint concrete.

This finding illustrates how the capabilities embodied in the four design requirements came together in practice: the ability to quickly try ideas at realistic scale, discuss observations with colleagues, and inspect results at the agent level surfaced planning considerations that abstract discussion alone had not produced.
Interactive simulation systems can contribute not only by evaluating candidate plans, but by making previously unarticulated planning vulnerabilities and questions visible.

\subsubsection{Reasoning Centered on Agent-Scale Inspection Over Crowd-Scale Aggregations} 

Participants consistently grounded their reasoning in specific, inspectable cases rather than relying primarily on aggregate metrics.
When outcomes looked surprising or concerning, they turned first to agent inspections, interviews, and localized crowd behavior to understand what was happening and why.
P1 used an analogy: \pquote{You can see that a team lost, but understanding \emph{why} requires watching the plays.}
P4, when interviewing agents who ran toward the shooter, described understanding their perspective even while disagreeing---the agent was prioritizing finding a friend, a choice she would not make but could understand given the agent's persona.
This finding suggests that the interview feature prompted users to engage with the agents' perspective rather than just to retrieve information alone.
 
We also found that evidence-based narratives about specific agent behaviors would be more persuasive than abstract statistics in organizational discussions.
Across individual sessions, 32 agent inspections and 3 interviews were logged, with additional inspections during the collaborative session.
This grounding also shaped how participants envisioned communicating with stakeholders.

This finding points toward a broader design insight: \emph{when simulations are used as reasoning environments, interpretability features serve not primarily for model validation but for evidence construction}---enabling experts to build arguments, identify edge cases, and communicate findings to others.

\subsubsection{Collaboration Centered on Shared Deliberation and Communal Sensemaking} 

The collaborative session followed participants’ prior individual exploration, allowing them to compare independently discovered insights and integrate them into shared planning decisions.
During this session, we found that the system’s collaborative value did not primarily come from simultaneous multi-user control, even though \textsc{WhatIf} supported it.
Rather, collaboration centered on using the simulation as a shared deliberation artifact for communal sensemaking: a common, inspectable reference point around which participants could point, question, compare, and revise ideas together.
P1 naturally took the lead on system interaction while P2 and P3 contributed verbally---a pattern consistent with how emergency planning teams already coordinate.
P2 confirmed that simultaneous multi-user input would be more useful in remote settings, but that in-person, single-operator control with group discussion made sense.
 
Even before touching the interface, the group debated accessibility, exit placement, police staging, and announcement strategy.
The collaborative run was the longest in the study (185 rounds) and produced one of the strongest outcomes, with the fewest fatalities and the most efficient evacuation.
These findings suggest that the shared simulation supported not only joint exploration and comparison of alternatives, but also the integration of independently developed insights into collective decision-making.

\subsubsection{Design Tensions and Failure Modes} \label{subsubsec:design_tensions}

Alongside these productive qualities, the evaluation surfaced important design tensions and failure modes for \textsc{WhatIf}.
We report these in detail because the way experts identified and responded to failures was itself analytically revealing, showing how they engaged with the system and what they expected from it.

\textit{Behavioral fidelity under extreme stress.}
The most salient failure occurred in the active shooter scenario, where agents congregated in orderly formations rather than dispersing chaotically.
P3 flagged this behavioral pattern as inconsistent with real-world panic behavior, and she then identified the root cause by interviewing the agents: agents were deliberate to achieve group consensus prior to moving, something that would be unlikely to occur under extreme threat.
Rather than dismissing the system, participants used this disagreement to articulate specific behavioral knowledge about instinct-driven evacuation and to identify what aspects of behavior the simulation would need to better capture.
 
\textit{Agent response homogeneity.}
P2 noted that agent interview responses were too similar, questioning whether the 12,000 agents represented meaningful behavioral diversity.
That she noticed this homogeneity indicates that she was inspecting agents with the expectation of finding meaningful variation---engaging critically rather than treating the simulation as an opaque tool.
The observation highlights a tension: the interpretability features that make the system useful depend on the underlying agent population exhibiting sufficient heterogeneity.

\textit{Tension between action and reflection.}
P4 noted that because \textsc{WhatIf} was so feature-rich, they sometimes forgot about the inspection capabilities it afforded.
The intervene--simulate--inspect--fork pattern observed in fork episodes indicates that reflection does occur, but may require deliberate pauses in the action cycle.
This finding suggests a design challenge: the system must make inspection feel like a natural part of the exploratory workflow rather than a separate analytical mode.

\subsubsection{Summary}

Taken together, these findings suggest that participants used \textsc{WhatIf} not simply to test predefined plans, but to explore alternatives, surface tacit assumptions, generate new planning considerations, and ground discussion in inspectable behavioral evidence.
In this sense, the LLM-powered social simulation functioned as a shared reasoning environment, supporting policy reasoning as an iterative, collaborative, and evidence-driven practice rather than a linear evaluation process.

Questionnaire responses broadly aligned with these qualitative findings: usability ratings were moderate to positive (mean SUS = 70.7), workload was primarily cognitive rather than physical, and participants reported particularly strong value for collaborative exploration, while responses related to agent-level interpretability were somewhat more variable. See Appendix Section~\ref{app:questionnaires} for detailed questionnaires and their respective scores.

More broadly, these results point to a design opportunity for simulation systems: supporting policymaking not only by producing outcomes, but by structuring how experts reason through uncertain futures together.

\section{Discussion}

Our findings suggest that the primary limitation of existing tools for policy planning under deep uncertainty is not only predictive accuracy, but the lack of support for interactive, collaborative reasoning.
Participants used \textsc{WhatIf} to iteratively explore alternatives, interrogate unexpected behaviors, and construct explanations together --- \textit{not} to identify one optimal plan.

\subsection{Simulations as Interactive Reasoning Environments}
 
A central implication of our evaluation is that the value of LLM-powered social simulations for expert use does not lie solely in improving predictive accuracy.
Instead, their value may lie equally in how they support the reasoning practices that experts already rely on under deep uncertainty: externalizing assumptions, probing alternatives, comparing consequences, and revising plans in response to emerging evidence. 
We position \textsc{WhatIf} to support scenario exploration, assumption surfacing, comparison, and critique; it is not intended to produce calibrated forecasts or final policy recommendations, and its outputs should not be interpreted as such.

This reframing has design consequences.
Traditional simulation systems optimize for output fidelity: the goal is to produce results that match the real world as closely as possible.
Our findings suggest that for interactive what-if exploration, the central design problem is how to make simulations usable for iterative reasoning.
The observation that the current exit plan can be inaccessible for wheelchair users emerged not from a high-fidelity behavioral reproduction but from a simulation that was good enough to make this planning gap visible.
P1's realization that his own expectations constituted a planning bias was triggered by agent behavior that surprised him, regardless of whether that behavior was perfectly calibrated.

Note that behavioral fidelity is still important---in our evaluation, we found that implausible agent behaviors undermined participant trust in simulation outcomes for specific scenarios.
But the productive use experts made of simulation failures suggests that the relationship between fidelity and utility is nonlinear.
Interactive LLM-powered social simulations occupy a design space between predictive models and structured thought experiments.
Within this space, usefulness depends not only on predictive fidelity, but on whether the system helps experts branch, inspect, and compare plausible consequences as their understanding evolves.

\subsection{Inspectability as a Mechanism for Surfacing Tacit Knowledge}

A recurring pattern in our evaluation was that simulation failures, when inspectable, became analytically productive rather than simply problematic.
This suggests that one of the most important interaction challenges in simulation-based decision support is not only helping experts see what happened, but helping them articulate why a simulated outcome does or does not align with their domain understanding.

This pattern suggests that simulation systems should make it easy for experts to articulate \emph{why} a simulation is wrong, not just \emph{that} it is wrong.
The combination of agent-level rationales and in-character interviews enabled participants to trace unexpected outcomes back to underlying behavioral assumptions, turning discrepancies into opportunities for analysis rather than sources of confusion.

More broadly, this finding positions simulations as productive boundary objects for expert reasoning~\cite{isenberg2011co,heer2007voyagers}.
Their value does not lie solely in being correct, but in being critique-able: providing a shared representation that experts can inspect, question, and refine.
LLM-powered social simulations,  which can produce expressive but occasionally implausible behaviors, may be well-suited to this role---provided that interfaces support critical inspection and explanation.

This perspective also reframes how trust should be designed for such systems~\cite{amershi2019guidelines,ehsan2023charting}.
Rather than treating trust as a static property established through validation benchmarks, our findings suggest that trust emerges through iterative interaction between experts and the model.
Participants engaged productively even when fidelity was imperfect, distinguishing between whether agent behavior appeared plausible and whether its underlying assumptions were sound.
This productive engagement implies that simulation systems should make disagreement actionable---for example, through making assumptions transparent, traceable decision processes, and structured pathways for expert correction.

\subsection{Simulations as Shared Deliberation Artifacts}
 
A key finding from the collaborative session was that the simulation’s value for teamwork did not primarily come from enabling simultaneous multi-user control.
Although \textsc{WhatIf} supported concurrent interaction, participants mainly used it as a shared deliberation artifact: one person operated the interface while others pointed, questioned, compared, and interpreted outcomes together.

Participants treated \textsc{WhatIf} as a common reference point that externalized the group's evolving hypotheses and made the consequences of proposed interventions immediately visible to all stakeholders.
Rather than concurrent editing, collaboration centered on integrating individually gathered findings and interpreting them together: participants first explored scenarios independently (T2), then used the shared simulation to compare alternatives, reconcile differing observations, and interpret implications across organizational roles (T3).

This finding suggests that collaborative simulation systems should be designed less around concurrent control and more around supporting discussion, pointing, comparison, and collective interpretation.
The system functioned as a shared deliberation artifact---a common reference point that externalized the group's evolving hypotheses and made the consequences of proposed interventions immediately visible to all stakeholders.
This finding extends prior work on collaborative visual analytics~\cite{viegas2007manyeyes,isenberg2011co} to a setting where the shared artifact is a dynamic, steerable simulation that responds to the group's evolving ideas.

Synchronized state, live presence, annotations, and cross-run comparison thus functioned less as tools for simultaneous control and more as mechanisms for grounding collective interpretation in a common, inspectable reference point.
We also note that P2 expected simultaneous multi-user input to be more useful in remote settings, whereas single-operator control with group discussion was preferred in the in-person, synchronous setting we explored here; the design implications of collaboration may therefore differ between co-located and distributed planning.
More broadly, collaborative simulation systems should support not only co-manipulation, but also the conversational and interpretive work through which teams make sense of uncertain futures together \cite{maitlis2005social}. 
 
\subsection{Limitations and Future Work} \label{subsec:lim}
 
This work also points to several open challenges in designing and evaluating simulation systems as reasoning environments for expert use.

For example, all five participants came from one organization, which limits generalizability.
P1--P3 had prior exposure through the formative study, which may have increased comfort with the system, but also enabled deeper expert engagement.
That participants belonged to a single team is both a strength---the evaluation captures a realistic collaborative context with shared institutional knowledge---and a limitation, as we cannot assess how the system would function across organizational boundaries with different planning cultures.
Scaling the evaluation to multiple organizations, planning domains, and scenario types would test the generality of the interaction patterns and design insights we observed.

Our evaluation did not include a baseline or comparison condition (e.g., tabletop-only planning).
Accordingly, our study could not isolate which benefits stem from \textsc{WhatIf}'s specific design versus interactive simulation more generally, and we present our findings as an exploratory account of how an expert team reasons with this paradigm rather than as a controlled comparison.
 
\textsc{WhatIf} also exposes several design tensions for future work.
The agent homogeneity concern points to a need for richer persona generation and prompting strategies that produce greater behavioral variation under stress~\cite{li2025actions}.
The tension between action and reflection suggests designs could explore contextual nudges that surface inspection capabilities during natural pauses.
Environmental fidelity gaps identified by participants (Appendix~\ref{sec:fidelity_gaps}) placed a ceiling on planning reasoning.
Selective deliberation and the deterministic spatial engine were necessary to sustain interactivity at this scale, but they involve a tradeoff: reducing per-agent LLM reasoning can lower behavioral richness, particularly under extreme stress (Section~\ref{subsubsec:design_tensions}).
Future work can examine this tradeoff more closely to identify the potential sweet spot between scale and behavioral richness for LLM-powered social simulations across different use cases.
Furthermore, structured validation of the simulation---scenario-level checks, sensitivity analysis of interventions, and tests of behavioral variation---remains important future work for establishing the boundaries within which its outputs can be trusted.

\section{Conclusion}

We present \textsc{WhatIf}, an interactive system that enables domain experts to collaboratively steer and inspect large-scale LLM-powered social simulations to support evidence-based policy reasoning.
Grou\-nded in a formative design engagement with emergency preparedness professionals and an evaluation across three evacuation scenarios, our work identifies four design requirements for interactive what-if exploration---fluid steering, real-time scale, collaborative exploration, and multi-level interpretability.
More broadly, we found that large-scale social simulations become most useful not as offline, individual-use, predictive tools, but as interactive, shared reasoning environments.
Their value extends well beyond predicting one optimal outcome: rather, they help experts iteratively compare and contrast uncertain futures as they reason through prospective policy decisions.

\begin{acks}
We thank I. Krsek for her valuable help with figure design and creation. This work was supported by the NOMIS foundation and the National Science Foundation under grant \#2316768.
\end{acks}

\bibliographystyle{ACM-Reference-Format}
\bibliography{sample-base}

\appendix
\setcounter{table}{0}
\setcounter{figure}{0}
\renewcommand{\thetable}{App.\arabic{table}}
\renewcommand{\thefigure}{App.\arabic{figure}}

\section{Representative Quotes from the Formative Study} \label{app:formative}

\subsection{DR1: Fluid Steering}

On how differently situated recipients interpret the same safety alert, one team member recalled:
\pquote{There's a demonstration, please avoid the area. You are going to tell me that I can go there. It's my freedom spirit.}

Another team member described the unintended effects of directive language:
\pquote{When we added the order, that basically created a [reaction]\ldots\ it was the blowback we got\ldots\ it was ideological.}

During a walkthrough, one team member spontaneously proposed testing a softer variant:
\pquote{I think it would be an interesting test to use the word asked\ldots\ if you just say people are being asked to avoid the area, I think people are going to treat that very differently, depending on their point of view.}

\subsection{DR2: Real-Time Scale}

On the limits of current planning exercises, one team member noted:
\pquote{We need data. And I think a lot of the tabletops and we do a lot of exercises, we have a lot of conversations, but we need data and we need to be able to analyze and make decisions with data and we usually don't create that much data [with] so many variables.}

The same team member contrasted typical coordination meetings with the questions he actually cares about:
\pquote{We usually have a meeting with police. We have the map. How would you respond? I will have three officers coming here\ldots\ But we don't go into the data. How long is it going to take? What are the other agents doing? How do they react?}

He also noted that conventional exercises flatten away human variation:
\pquote{When you normally do tabletops or exercises, it's a blank human. You don't know who they are. They're all going to act the same, but that's not true in real life scenarios.}

On the burden of manual monitoring, one team member described:
\pquote{If I log in in the morning and I see 257 results, I have to go through 257 results to make sure there's nothing here that is of danger or is unimportant.}

The same team member observed that existing tools lack forward-looking insight:
\pquote{The insight and the summary of---we need to be watching for more protests like this in the future---that's something none of these pieces of software can give us.}

On family dynamics during evacuation, one team member reasoned:
\pquote{For every student, there is potentially minimum of two family members\ldots\ there is a feeling associated with where exactly are they and do I go to them and help them evacuate versus going in the direction that I need to.}

\subsection{DR3: Collaborative Exploration}

On fragmented visibility across teams, one team member stated:
\pquote{We don't always know what the other groups are working on. I would like it to be that way---so we can all be on the same page at one time---but it doesn't always work out that way.}

\subsection{DR4: Multi-Level Interpretability}

On treating simulation as a diagnostic tool rather than a black-box predictor, one team member explained:
\pquote{We see this as a learning tool\ldots\ every time that we review, there's always something we're going to be learning something new\ldots\ we do trust these to give us information to know where do we need to fill gaps.}

\section{Additional Information on the User Study}

\subsection{Questionnaires Used and Their Scores} \label{app:questionnaires}
\begin{table}[ht]
\centering
\small
\caption{NASA-TLX items (7-point scale).}
\label{tab:nasatlx}
\begin{tabular}{c p{0.82\columnwidth}}
\toprule
\textbf{\#} & \textbf{Question} \\
\midrule
N1 & How mentally demanding was the task? \\
N2 & How physically demanding was the task? \\
N3 & How hurried or rushed was the pace of the task? \\
N4 & How successful were you in accomplishing what you were asked to do? \\
N5 & How hard did you have to work to accomplish your level of performance? \\
N6 & How insecure, discouraged, irritated, stressed, and annoyed were you? \\
\bottomrule
\end{tabular}
\end{table}

\begin{figure}[ht]
    \centering
    \includegraphics[width=\linewidth]{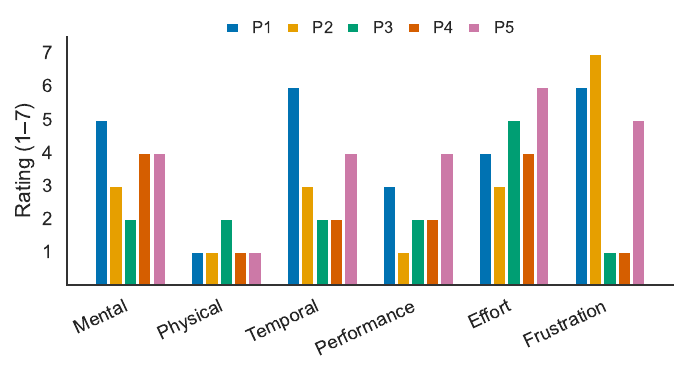}
    \caption{This figure shows NASA-TLX workload ratings across dimensions. Workload is characterized by moderate mental and temporal demand, low physical demand, and varied frustration across participants.}
    \Description{Grouped bar chart of NASA-TLX dimensions (Mental, Physical, Temporal, Performance, Effort, Frustration) rated 1--7 for five participants, showing higher mental/temporal demand and varied frustration.}
    \label{fig:app_nasa}
\end{figure}

\begin{table}[ht]
\centering
\small
\caption{System Usability Scale (SUS) items (7-point Likert).}
\label{tab:sus}
\begin{tabular}{c p{0.82\columnwidth}}
\toprule
\textbf{\#} & \textbf{Statement} \\
\midrule
S1  & I would like to use this system frequently. \\
S2  & I found the system unnecessarily complex. \\
S3  & I thought the system was easy to use. \\
S4  & I would need technical support to use this system. \\
S5  & The various functions were well integrated. \\
S6  & There was too much inconsistency in this system. \\
S7  & Most people would learn to use this system quickly. \\
S8  & I found the system very cumbersome to use. \\
S9  & I felt very confident using the system. \\
S10 & I needed to learn a lot before I could get going. \\
\bottomrule
\end{tabular}
\end{table}

\begin{figure}[ht]
    \centering
    \includegraphics[width=\linewidth]{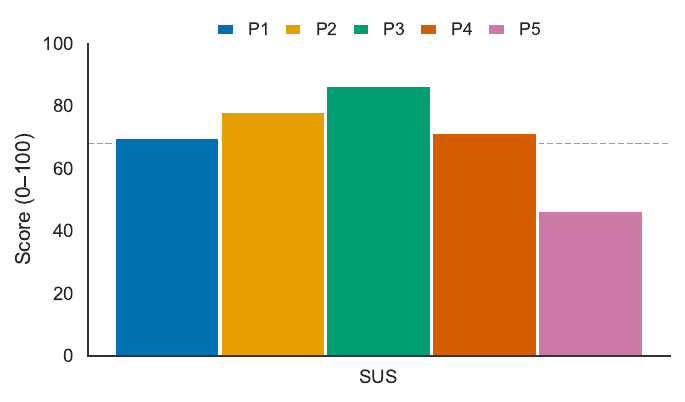}
    \caption{This figure shows System Usability Scale (SUS) scores for each participant. Participants reported generally high perceived usability of the WhatIf system, with most scores above the standard benchmark.}
    \Description{Bar chart of SUS scores (0--100) for five participants (P1--P5), with most scores above average, P3 highest, and P5 lower.}
    \label{fig:app_sus}
\end{figure}

\begin{table}[ht]
\centering
\small
\caption{Custom questionnaire (7-point Likert). $\dagger$~=~reverse-scored.}
\label{tab:custom}
\begin{tabular}{l c p{0.62\columnwidth}}
\toprule
\textbf{Category} & \textbf{\#} & \textbf{Statement} \\
\midrule
\multirow{5}{*}{\parbox{1.4cm}{\scriptsize\centering \textbf{DR1}\\\textbf{Fluid}\\  \textbf{steering}}}
 & C1  & I could intervene in the simulation quickly. \\
 & C2  & The system gave me enough ways to influence agent behavior. \\
 & C3  & The What-If Agent was a useful way to author interventions. \\
 & C4  & I received immediate visual feedback after intervening. \\
 & C5$^\dagger$ & The simulation didn't feel responsive. \\
\midrule
\parbox{1.4cm}{\scriptsize\centering \textbf{DR2}\\\textbf{Real-time}\\\textbf{scale}}
 & C6  & I could understand what was happening despite 12,000+ agents. \\
\midrule
\multirow{3}{*}{\parbox{1.4cm}{\scriptsize\centering \textbf{DR3}\\\textbf{Collab.}\\\textbf{exploration}}}
 & C7$^\dagger$ & I was not aware of what my collaborators were doing. \\
 & C8  & Shared cursors, annotations, and Overhear helped us coordinate. \\
 & C9  & Working as a team produced a better outcome than individually. \\
\midrule
\multirow{5}{*}{\parbox{1.4cm}{\scriptsize\centering \textbf{DR4}\\\textbf{Multi-level}\\\textbf{interpret.}}}
 & C10  & I trusted that agents' decisions were plausible. \\
 & C11$^\dagger$ & I could not understand why individual agents made specific decisions. \\
 & C12 & Rationale and interview features helped me understand agent reasoning. \\
 & C13 & Visualizations and comparison tools helped identify effective strategies. \\
 & C14$^\dagger$ & It's hard to compare strategies using the comparison and report features. \\
\midrule
\multirow{2}{*}{\parbox{1.4cm}{\scriptsize\centering \textbf{Adoption}}}
 & C15 & The system is a valuable addition to our preparedness process. \\
 & C16 & I would use this system in future preparedness processes. \\
\bottomrule
\end{tabular}
\end{table}

\begin{figure}[ht]
    \centering
    \includegraphics[width=\linewidth]{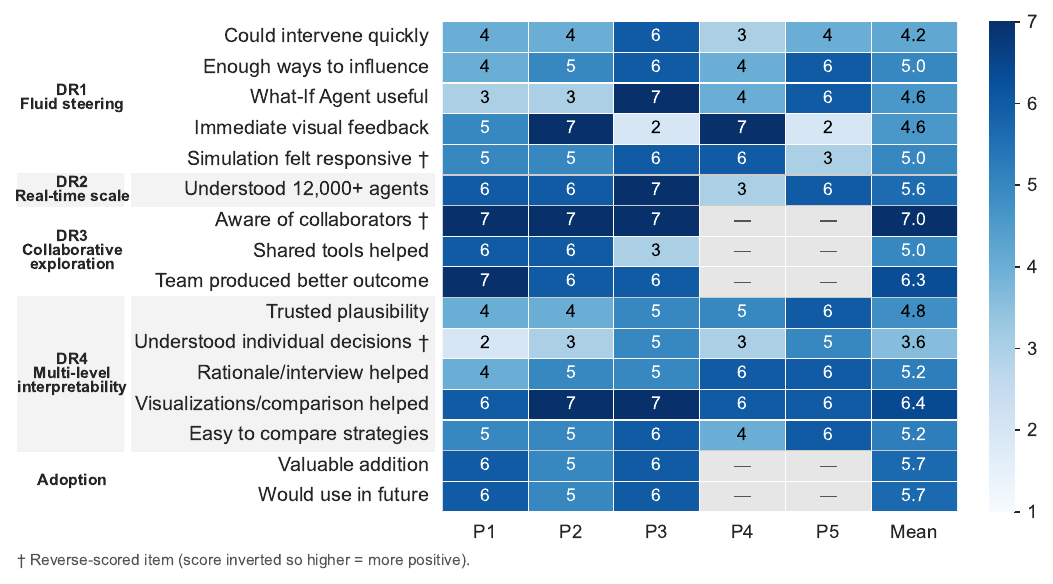}
    \caption{This figure shows Likert-scale ratings of system features grouped by design requirements. Participants rated WhatIf highly on collaboration, interpretability, and exploration features, particularly for awareness and comparative reasoning.}
    \Description{Heatmap of Likert-scale (1--7) responses across system features grouped by design requirements, with generally high ratings and darker cells indicating stronger agreement.}
    \label{fig:app_cus}
\end{figure}

\begin{table}[ht]
\centering
\footnotesize
\caption{Questionnaire score summary for NASA-TLX, SUS, and design-requirement means.}
\label{tab:whatif-survey-summary}
\begin{tabular}{lccccc>{\columncolor{gray!10}}c}
\toprule
Measure & P1 & P2 & P3 & P4 & P5 & Mean (SD) \\
\midrule
\multicolumn{7}{l}{\textbf{NASA-TLX}} \\
Mental Demand & 5 & 3 & 2 & 4 & 4 & 3.6 (1.1) \\
Physical Demand & 1 & 1 & 2 & 1 & 1 & 1.2 (0.4) \\
Temporal Demand & 6 & 3 & 2 & 2 & 4 & 3.4 (1.7) \\
Performance & 3 & 1 & 2 & 2 & 4 & 2.4 (1.1) \\
Effort & 4 & 3 & 5 & 4 & 6 & 4.4 (1.1) \\
Frustration & 6 & 7 & 1 & 1 & 5 & 4 (2.8) \\
\addlinespace[2pt]
\multicolumn{7}{l}{\textbf{SUS}} \\
SUS Score & 70 & 78.3 & 86.7 & 71.7 & 46.7 & 70.7 (14.9) \\
\addlinespace[2pt]
\multicolumn{7}{l}{\textbf{DR Means}} \\
DR1 Fluid steering & 4.2 & 4.8 & 5.4 & 4.8 & 4.2 & 4.7 (0.5) \\
DR2 Real-time scale & 6 & 6 & 7 & 3 & 6 & 5.6 (1.5) \\
DR3 Collaborative exploration & 6.7 & 6.3 & 5.3 & — & — & 6.1 (0.7) \\
DR4 Multi-level interpretability & 4.2 & 4.8 & 5.6 & 4.8 & 5.8 & 5.0 (0.7) \\
Adoption & 6 & 5 & 6 & — & — & 5.7 (0.6) \\
\bottomrule
\end{tabular}
\vspace{2pt}
\parbox{\linewidth}{\footnotesize Reversed custom items are inverted before averaging.}
\end{table}

\section{Time per Round}

\begin{figure}[ht]
    \centering
    \includegraphics[width=\linewidth]{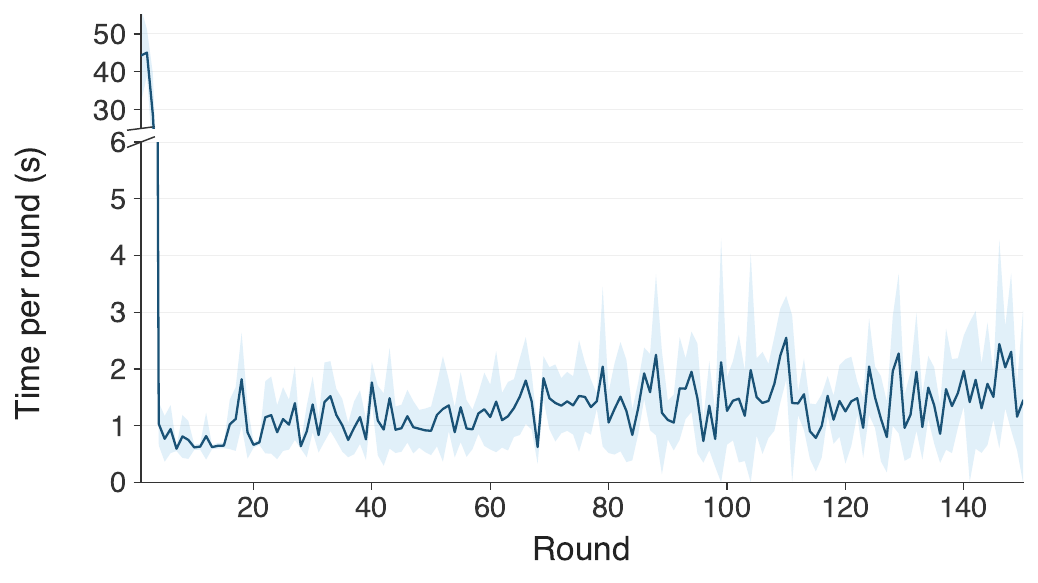}
    \caption{Mean time per simulation round across five runs (shaded region: ±1 SD). The first few rounds take significantly longer due to initialization overhead, after which computation stabilizes to approximately 1–2 seconds per round.}
    \Description{A line chart showing the time per simulation round (in seconds) over approximately 150 rounds. The x-axis represents the round number, and the y-axis represents time per round in seconds, with a broken axis indicating an early outlier. At the beginning (around rounds 1–3), there is a sharp spike in execution time, reaching over 40–50 seconds. Immediately after, the time drops steeply to below 2 seconds. For the remainder of the rounds, execution time stays relatively low, generally fluctuating between about 0.5 and 2 seconds, with occasional spikes reaching around 3–4 seconds. A shaded region around the line indicates variability, which increases slightly in later rounds. Overall, the chart shows that after an initial expensive startup phase, most simulation rounds complete quickly with modest variability.}
    \label{fig:round_time}
\end{figure}

\section{Environmental Fidelity Gaps}
\label{sec:fidelity_gaps}

Participants identified several environmental features not modeled in the current simulation that constrained planning reasoning:
entrances that may function as emergency exits, with configurable probabilities of use different from regular exits;
extended geography beyond the simulation boundary that limited reasoning about coordination between field and command;
and underground tunnels connecting parts of the venue that planners consider in practice but were absent from the model.
These gaps did not prevent productive use of the system but placed a ceiling on how far participants could push specific planning questions before encountering simulation limitations.

\definecolor{rowshade}{HTML}{F5F5F5}
\definecolor{hdrcolor}{HTML}{2C3E50}

\newcolumntype{L}[1]{>{\raggedright\arraybackslash}p{#1}}

\newcommand{\tref}[1]{\textsuperscript{\textit{#1}}}
\newcommand{\tnote}[2]{\textsuperscript{\textit{#1}}\,#2}

\section{Design-Choice Rationale} \label{app:design-choices}

The following tables document every modelling choice in the simulator together with the public source (or explicit acknowledgement of a calibration assumption) that motivates it.  Where a parameter is described as a \emph{surrogate} or \emph{simplification}, no claim of empirical fit is made; the basis column records the literature that bounds the choice.

\begin{table*}[t]
\caption{Design choices for \textbf{fire} threat modelling.}\label{tab:fire}
\centering
\begin{tabular}{%
    L{0.16\textwidth}
    L{0.38\textwidth}
    L{0.38\textwidth}
  }
\toprule
\rowcolor{hdrcolor}
\textcolor{white}{\textbf{Choice}} &
\textcolor{white}{\textbf{Description}} &
\textcolor{white}{\textbf{Basis}} \\
\midrule

\rowcolor{rowshade}
Severity-to-growth mapping &
Spreading fires: \texttt{low}\,=\,slow, \texttt{medium}\,=\,medium, \texttt{high}\,=\,ultra-fast. &
NIST $t$-squared design-fire classes reaching 1\,055\,kW in 600\,s, 300\,s, 150\,s, and 75\,s.\tref{a} \\

Radial growth model &
$t^{2}$ heat release yields approximately linear radial growth ($r\!\propto\!\sqrt{Q}\!\propto\! t$). &
Point-source radiation contour with $Q=\alpha t^{2}$.\tref{a,b} \\

\rowcolor{rowshade}
Radiative fraction &
Assumed $\chi_r = 0.35$ for contour growth. &
Common engineering default for flaming fires.\tref{b} \\

Tenability / injury flux bands &
2.5, 5, 10\,kW/m$^{2}$. &
NIST: 2.5\,kW/m$^{2}$ tenability threshold; 5\,kW/m$^{2}$ $\approx$\,30\,s to 2nd-degree burns; 10\,kW/m$^{2}$ $\approx$\,10\,s.\tref{b} \\

\rowcolor{rowshade}
Fatality onset rounds &
Mapped from NIST flux-exposure durations via ${\sim}$2--3\,s/round. &
NIST flux durations converted to simulator rounds.\tref{b} \\

Per-round fatality escalation &
Simplified hazard surrogate (no FED, suppression, or rescue modelling). &
NIST tenability data motivates thresholds; post-onset roll is an acknowledged simplification.\tref{b} \\

\bottomrule
\end{tabular}

\vspace{4pt}
{\raggedright\footnotesize
\tnote{a}{NIST TN 1889v2: \url{https://nvlpubs.nist.gov/nistpubs/TechnicalNotes/NIST.TN.1889v2.pdf}}\par
\tnote{b}{NIST Fire Dynamics Tools (NUREG-1805): \url{https://tsapps.nist.gov/publication/get_pdf.cfm?pub_id=911564}}\par
}
\end{table*}

\begin{table*}[t]
\caption{Design choices for \textbf{bomb\,/\,explosive} threat modelling.}\label{tab:bomb}
\centering
\begin{tabular}{%
    L{0.16\textwidth}
    L{0.38\textwidth}
    L{0.38\textwidth}
  }
\toprule
\rowcolor{hdrcolor}
\textcolor{white}{\textbf{Choice}} &
\textcolor{white}{\textbf{Description}} &
\textcolor{white}{\textbf{Basis}} \\
\midrule

\rowcolor{rowshade}
Severity-to-device mapping &
\texttt{low}\,=\,5\,lb pipe bomb; \texttt{medium}\,=\,20\,lb vest; \texttt{high}\,=\,50\,lb suitcase bomb. &
FEMA safe-distance tables.\tref{a} \\

Visible threat radius &
FEMA mandatory evacuation distances: 21, 34, 46\,m. &
Closest official stand-off values at venue scale.\tref{a} \\

\rowcolor{rowshade}
Lethal core radii &
8\,m (pipe bomb), 12\,m (suitcase bomb); medium interpolated via cube-root TNT scaling. &
FEMA lethal air-blast anchors with Hopkinson--Cranz scaling.\tref{a,b} \\

Consequence model &
Immediate lethal core $+$ crowd avoidance\,/\,stampede; no fuse, shielding, or fragment modelling. &
Conservative surrogate grounded in FEMA stand-off geometry.\tref{a,b} \\

\bottomrule
\end{tabular}

\vspace{4pt}
{\raggedright\footnotesize
\tnote{a}{FEMA 453: \url{https://www.fema.gov/sites/default/files/2020-08/fema453.pdf}}\par
\tnote{b}{FEMA IS-15: \url{https://emilms.fema.gov/IS15b/groups/215.html}}\par
}
\end{table*}

\begin{table*}[t]
\caption{Design choices for \textbf{shooter\,/\,police} modelling (Part~I --- weapon mechanics).}\label{tab:shooter1}
\centering
\begin{tabular}{%
    L{0.16\textwidth}
    L{0.38\textwidth}
    L{0.38\textwidth}
  }
\toprule
\rowcolor{hdrcolor}
\textcolor{white}{\textbf{Choice}} &
\textcolor{white}{\textbf{Description}} &
\textcolor{white}{\textbf{Basis}} \\
\midrule

\rowcolor{rowshade}
Magazine size &
30 rounds. &
Army rifle/carbine doctrine.\tref{a} \\

Firing cadence &
Rapid semi-auto $\approx$\,45\,rpm; split into stationary vs.\ moving fire per round. &
TC~3-22.9 rapid semi-auto rate, mapped to 2--3\,s rounds.\tref{a} \\

\rowcolor{rowshade}
Reload duration &
Conservative tactical pause for a magazine change under stress. &
No clean public doctrinal source; acknowledged calibration assumption. \\

Effective \& max range &
Compressed to crowd-interaction distances (``reliably hit a moving person in chaos''). &
Doctrinal max effective range as upper bound.\tref{a} \\

\rowcolor{rowshade}
Hit-probability curve &
Blend of stress-fire accuracy data and rifle controllability at short range. &
Public anchor: NYPD stress-fire data.\tref{b} \\

Lethality curve &
Higher close-range lethality for centre-mass rifle hits; not all GSWs fatal. &
Assault-firearm case-fatality data~\cite{kaufman2021epidemiologic}. \\

\rowcolor{rowshade}
Wound speed factor &
Mobility-impairment simplification; wounded agents meaningfully slower. &
Modelling assumption (no medical estimate claimed). \\

\bottomrule
\end{tabular}

\vspace{4pt}
{\raggedright\footnotesize
\tnote{a}{TC~3-22.9 Rifle and Carbine: \url{https://www.moore.army.mil/Infantry/199th/ocs/content/pdf/TC\%203-22.9\%20Rifle\%20and\%20Carbine.pdf}}\par
\tnote{b}{NYPD 2012 Firearms Discharge Report: \url{https://www.nyc.gov/assets/nypd/downloads/pdf/public_information/2012fdr.pdf}}\par
}
\end{table*}

\begin{table*}[t]
\caption{Design choices for \textbf{shooter\,/\,police} modelling (Part~II --- behaviour \& police response).}\label{tab:shooter2}
\centering
\begin{tabular}{%
    L{0.16\textwidth}
    L{0.38\textwidth}
    L{0.38\textwidth}
  }
\toprule
\rowcolor{hdrcolor}
\textcolor{white}{\textbf{Choice}} &
\textcolor{white}{\textbf{Description}} &
\textcolor{white}{\textbf{Basis}} \\
\midrule

\rowcolor{rowshade}
Shooter state machine &
Stationary fire\,/\,walking fire\,/\,running\,/\,reloading. &
FBI\tref{a} and PERF\tref{b} active-shooter reporting. \\

State durations \& reposition trigger &
Tactical pacing to avoid static-turret artefact; forces re-targeting. &
No primary per-state dwell-time dataset; acknowledged assumption. \\

\rowcolor{rowshade}
Movement randomness &
Perturbation to avoid deterministic pursuit. &
Not calibrated to human path variability. \\

Officer--shooter duel probabilities &
Calibrated from FBI event-level outcomes with LE gunfire exchange. &
FBI 2000--2013 study.\tref{a} \\

\rowcolor{rowshade}
Officer speed &
Conservative purposeful movement with gear, below full sprint. &
No transferable public dataset; acknowledged assumption. \\

Engage radius &
Close to the 21-ft\,/\,6.4\,m officer-safety distance. &
Classic lethal-threat training distance; close-contact proxy. \\

\rowcolor{rowshade}
Suppression factor &
Disruption and hesitation once officers close distance. &
No public primary source for radius$+$multiplier pair. \\

\bottomrule
\end{tabular}

\vspace{4pt}
{\raggedright\footnotesize
\tnote{a}{FBI Active Shooter Study 2000--2013: \url{https://www.fbi.gov/file-repository/reports-and-publications/active-shooter-study-2000-2013-1.pdf/view}}\par
\tnote{b}{PERF Active Shooter Report: \url{https://www.policeforum.org/assets/ActiveShooter.pdf}}\par
}
\end{table*}

\begin{table*}[t]
\caption{Design choices for \textbf{severe weather} (lightning) modelling.}\label{tab:weather}
\centering
\begin{tabular}{%
    L{0.16\textwidth}
    L{0.38\textwidth}
    L{0.38\textwidth}
  }
\toprule
\rowcolor{hdrcolor}
\textcolor{white}{\textbf{Choice}} &
\textcolor{white}{\textbf{Description}} &
\textcolor{white}{\textbf{Basis}} \\
\midrule

\rowcolor{rowshade}
Lightning as primary threat &
Open-air stadiums explicitly vulnerable; sheltering is the required protective action. &
NWS lightning safety\tref{a} and organised-sports guidance.\tref{b} \\

Strike-cadence mapping &
\texttt{low}\,=\,1--3, \texttt{med}\,=\,4--11, \texttt{high}\,=\,12+ flashes/min; Poisson sampling per 2.5\,s round. &
NWS cloud-to-ground flash-rate categories.\tref{c} \\

\rowcolor{rowshade}
Casualty zone radius &
${\sim}$100\,ft\,/\,30.5\,m ground-current radius per strike. &
NOAA: injuries reported up to ${\sim}$100\,ft;\tref{d} ground current is the primary casualty mechanism.\tref{e} \\

Fatal\,/\,nonfatal split &
``Almost 90\,\% survive'' applied per victim in the strike zone. &
CDC national survival fraction.\tref{f} \\

\bottomrule
\end{tabular}

\vspace{4pt}
{\raggedright\footnotesize
\tnote{a}{NWS Lightning Safety: \url{https://www.weather.gov/safety/lightning}}\par
\tnote{b}{NWS Organised Sports \& Lightning: \url{https://www.weather.gov/media/safety/OrganizedOutdoorSportsActivitiesandLightning.pdf}}\par
\tnote{c}{NWS Melbourne Lightning Threat Definitions: \url{https://www.weather.gov/mlb/lightning_threat}}\par
\tnote{d}{NOAA Backcountry Lightning Guide: \url{https://www.weather.gov/media/safety/backcountry_lightning.pdf}}\par
\tnote{e}{NWS Ground-Current Science: \url{https://www.weather.gov/safety/lightning-science-ground-currents}}\par
\tnote{f}{CDC Lightning Victim Data: \url{https://www.cdc.gov/lightning/data-research/index.html}}\par
}
\end{table*}

\begin{table*}[t]
\caption{Design choices for \textbf{hazardous materials} (chlorine release) modelling.}\label{tab:hazmat}
\centering
\begin{tabular}{%
    L{0.16\textwidth}
    L{0.38\textwidth}
    L{0.38\textwidth}
  }
\toprule
\rowcolor{hdrcolor}
\textcolor{white}{\textbf{Choice}} &
\textcolor{white}{\textbf{Description}} &
\textcolor{white}{\textbf{Basis}} \\
\midrule

\rowcolor{rowshade}
Chlorine as threat agent &
Common inhalation hazard with public AEGL, IDLH, and emergency-response guidance. &
NIOSH emergency response card\tref{a} and Pocket Guide.\tref{b} \\

Threat radius $=$ IDLH contour &
Outer circle represents the 10\,ppm NIOSH IDLH boundary. &
NIOSH IDLH documentation.\tref{c} \\

\rowcolor{rowshade}
Inner contour dilution &
Inverse-square dilution anchored at the IDLH boundary. &
Explicit simplification preserving circle mechanic.\tref{a,c} \\

Concentration injury bands &
30\,ppm (intense coughing), 40--60\,ppm (pulmonary oedema), 330\,ppm\,/\,5\,min (acute lethality). &
NIOSH IDLH derivation\tref{c} and ATSDR toxicological profile.\tref{d} \\

\rowcolor{rowshade}
Fatality onset rounds &
5\,min and 30\,min exposure durations $\rightarrow$ round counts at 2.5\,s/round. &
Published durations converted to simulator time.\tref{a,c} \\

Post-onset fatality escalation &
Surrogate hazard roll; no wind, CFD, dose integration, or decon modelling. &
Official thresholds ground onset; post-onset acknowledged as simplification.\tref{a,d} \\

\bottomrule
\end{tabular}

\vspace{4pt}
{\raggedright\footnotesize
\tnote{a}{NIOSH Chlorine Emergency Response Card: \url{https://www.cdc.gov/niosh/ershdb/emergencyresponsecard_29750024.html}}\par
\tnote{b}{NIOSH Pocket Guide --- Chlorine: \url{https://www.cdc.gov/niosh/npg/npgd0115.html}}\par
\tnote{c}{NIOSH IDLH --- Chlorine: \url{https://www.cdc.gov/niosh/idlh/7782505.html}}\par
\tnote{d}{ATSDR Toxicological Profile for Chlorine: \url{https://www.atsdr.cdc.gov/toxprofiles/tp172.pdf}}\par
}
\end{table*}

\begin{table*}[t]
\caption{Design choices for \textbf{crowd dynamics and navigation} (Part~I --- motion models \& stampede).}\label{tab:crowd1}
\centering
\begin{tabular}{%
    L{0.16\textwidth}
    L{0.38\textwidth}
    L{0.38\textwidth}
  }
\toprule
\rowcolor{hdrcolor}
\textcolor{white}{\textbf{Choice}} &
\textcolor{white}{\textbf{Description}} &
\textcolor{white}{\textbf{Basis}} \\
\midrule

\rowcolor{rowshade}
Social-force parameters &
$\tau$, $A$, $B$, $\lambda$, $k_{\mathrm{body}}$, $\kappa_{\mathrm{friction}}$. &
Helbing \& Moln\'{a}r~\cite{helbing1995social} and Helbing et al.~\cite{helbing2000simulating}. \\

Fundamental-diagram parameters &
$v_{\mathrm{free}}$, $\rho_{\mathrm{crit}}$, $\rho_{\mathrm{jam}}$, $\gamma$. &
Weidmann~\cite{weidmann1993transporttechnik} and Seyfried et al.~\cite{seyfried2005fundamental}. \\

\rowcolor{rowshade}
ORCA parameters &
$\tau$, $\tau_{\mathrm{obst}}$, max neighbours, neighbour distance. &
Standard practical defaults from ORCA\,/\,RVO2~\cite{van2011reciprocal}.\tref{a} \\

Fruin-based crowd slowdown &
Slowdown onset $\approx$\,25\,sq\,ft/person; bottoms at $\approx$\,5\,sq\,ft/person. &
Fruin's LOS regimes~\cite{fruin1970designing}. \\

\rowcolor{rowshade}
Stampede trigger density &
Onset near Fruin LOS\,F (${\sim}$3.6\,p/m$^{2}$); saturates at 6\,p/m$^{2}$. &
Fruin~\cite{fruin1970designing} and Helbing et al.~\cite{helbing2007dynamics}. \\

Stampede probability \& cap &
Calibrated hazard knobs; not measured event frequencies. &
Crowd turbulence literature~\cite{golas2014continuum}. \\

\rowcolor{rowshade}
Stampede severity multipliers &
Maps threat seriousness to panic pressure. &
Modelling choice (no direct psychometric fit). \\

\bottomrule
\end{tabular}

\vspace{4pt}
{\raggedright\footnotesize
\tnote{a}{RVO2 library documentation: \url{https://gamma-web.iacs.umd.edu/RVO2/documentation/2.0/}}\par
}
\end{table*}

\begin{table*}[t]
\caption{Design choices for \textbf{crowd dynamics and navigation} (Part~II --- hazards, perception \& layout).}\label{tab:crowd2}
\centering
\begin{tabular}{%
    L{0.16\textwidth}
    L{0.38\textwidth}
    L{0.38\textwidth}
  }
\toprule
\rowcolor{hdrcolor}
\textcolor{white}{\textbf{Choice}} &
\textcolor{white}{\textbf{Description}} &
\textcolor{white}{\textbf{Basis}} \\
\midrule

\rowcolor{rowshade}
Stampede cascade multiplier &
Shock-wave and turbulence propagation ideas. &
Helbing et al.~\cite{helbing2007dynamics} and Still~\cite{still2014introduction}. \\

Stampede duration \& fatality prob. &
Simplified immobilisation and crush risk. &
Prolonged-immobilisation literature.\tref{a} \\

\rowcolor{rowshade}
Threat-awareness zone &
$1.5\times$ threat radius as a cue-perception buffer. &
NIST protective-action behaviour models.\tref{b,c} \\

Local congestion factor &
Numerical stabiliser for tight bottlenecks. &
Bottleneck flow empirics~\cite{adrian2020crowds}. \\

\rowcolor{rowshade}
Path-cost hierarchy &
People prefer aisles over crossing seat rows. &
Tacit knowledge and policymaker experience. \\

Visual surroundings radius &
Local person/feature descriptions at Hall personal-space scale. &
Hall proxemics~\cite{hall1966hidden}. \\

\rowcolor{rowshade}
Coordinator influence radius &
Staff-member ``close enough to notice and hear'' zone. &
Hall social-distance regime~\cite{hall1966hidden}. \\

\bottomrule
\end{tabular}

\vspace{4pt}
{\raggedright\footnotesize
\tnote{a}{UpToDate --- Severe Crush Injury in Adults: \url{https://www.uptodate.com/contents/severe-crush-injury-in-adults}}\par
\tnote{b}{NIST TN 2191: \url{https://nvlpubs.nist.gov/nistpubs/TechnicalNotes/NIST.TN.2191.pdf}}\par
\tnote{c}{NIST TN 1827: \url{https://nvlpubs.nist.gov/nistpubs/TechnicalNotes/NIST.TN.1827.pdf}}\par
}
\end{table*}

\begin{table*}[t]
\caption{Design choices for \textbf{coordination and active controls}.}\label{tab:controls}
\centering
\begin{tabular}{%
    L{0.16\textwidth}
    L{0.38\textwidth}
    L{0.38\textwidth}
  }
\toprule
\rowcolor{hdrcolor}
\textcolor{white}{\textbf{Choice}} &
\textcolor{white}{\textbf{Description}} &
\textcolor{white}{\textbf{Basis}} \\
\midrule

\rowcolor{rowshade}
Coordinator reaction probability &
$p=0.5$; moderate-following assumption. &
NIST protective-action and evacuation behaviour reviews.\tref{a,b} \\

Discussion cap &
\texttt{max\_discussion\_rounds}\,=\,3; computational cap balancing LLM latency and behavioural estimates. &
Real-world pre-evacuation delay ranges.\tref{c} \\

\bottomrule
\end{tabular}

\vspace{4pt}
{\raggedright\footnotesize
\tnote{a}{NIST TN 2191: \url{https://nvlpubs.nist.gov/nistpubs/TechnicalNotes/NIST.TN.2191.pdf}}\par
\tnote{b}{NIST TN 1827: \url{https://nvlpubs.nist.gov/nistpubs/TechnicalNotes/NIST.TN.1827.pdf}}\par
\tnote{c}{NIST TN 1839: \url{https://nvlpubs.nist.gov/nistpubs/TechnicalNotes/NIST.TN.1839.pdf}}\par
}
\end{table*}

\clearpage

\renewcommand{\algorithmicrequire}{\textbf{Input:}}
\renewcommand{\algorithmicensure}{\textbf{Output:}}
\renewcommand{\algorithmiccomment}[1]{\hfill$\triangleright$ #1}

\newenvironment{whatifimplalgorithm}[1]{%
  \begin{algorithm}[H]
  \caption{#1}
  \small
  \begin{algorithmic}[1]
}{%
  \end{algorithmic}
  \end{algorithm}
}

\section{System Implementation Details} \label{app:implementation}

WhatIf follows the same architectural separation described in the main paper: an LLM decision layer determines where agents want to go and why, while a deterministic spatial layer executes movement, hazard evolution, and crowd-level effects. Concretely, the runtime stack consists of a Python backend built on FastAPI, async I/O, and SQLite, plus a React frontend connected through project-scoped WebSockets. The appendix below focuses on the implementation mechanisms that make the system interactive at scale: exact state restoration, selective batched deliberation, a unified intervention engine, deterministic movement and hazard updates, shared multi-user synchronization, and snapshot-based replay and recap generation.

\subsection{Runtime State and Exact Restoration}
Each live run is represented as a \texttt{SimulationInstance} containing the 12{,}278-agent population, an \texttt{AgentIndex} for constant-time lookup, a \texttt{GroupDestinationTracker}, cached coordinator influence zones, the current condition, and any environment overlays such as threats, custom exits, obstacles, or stage offsets. A notable implementation detail is that some high-performance subsystems, including navigation grids, exit registries, and coordinator caches, keep mutable module-level state. Accordingly, \texttt{SimulationManager} treats initialization, loading, and simulation switching as explicit snapshot/restore operations: a fresh run resets those globals, loads the pre-generated population and prompt assets, and instantiates agents with persona-conditioned prompts, whereas a restored run first recreates the same baseline agents and then overwrites their positions, states, targets, histories, deliberation metadata, and group-tracker maps from saved runtime state when available. This is what makes pausing, loading, and branching reliable in practice: each saved simulation can be resumed as a runnable stateful process rather than merely replayed as a frozen visualization. The population of 12{,}278 personas and their group memberships was generated by sampling demographic and role attributes from distributions derived from emergency preparedness professionals and grouping co-attending agents into families and social units. We do not guarantee behavioral diversity across this population; the response homogeneity participants observed under extreme stress (Section~\ref{subsubsec:design_tensions}) and the need for richer persona generation (Section~\ref{subsec:lim}) reflect this limitation.

\begin{whatifimplalgorithm}{Initializing or Restoring a Simulation Instance}
\REQUIRE condition index $c$, optional announcement override $a$, optional snapshot $S$, optional runtime payload $R$
\ENSURE live simulation instance $\mathcal{S}$
\STATE Load prompt assets, population records, initial positions, and condition templates if not cached
\STATE Reset mutable global environment state: coordinator positions, exit registry, navigation grid, and map caches
\STATE Instantiate agents from the pre-generated population using condition $c$ and announcement $a$
\IF{$S \neq \varnothing$}
  \STATE Overwrite coarse agent positions and states from snapshot $S$
  \STATE Restore environment overlays (coordinators, exits, obstacles, threats) from $S$ or $R$
\ENDIF
\STATE Build \textsc{AgentIndex}, \textsc{GroupDestinationTracker}, and \textsc{CoordinatorZones}
\IF{$R \neq \varnothing$}
  \FORALL{saved agent payloads in $R$}
    \STATE Restore targets, decision flags, histories, rationales, exposure counters, and round-local metadata
  \ENDFOR
  \STATE Restore tracker maps, simulation attributes, and the saved round number
\ENDIF
\STATE Snapshot the current mutable environment globals back into the instance
\STATE Register $\mathcal{S}$ under a new live simulation identifier and return it
\end{whatifimplalgorithm}

\subsection{Selective Batched Deliberation}
The LLM layer is invoked selectively rather than once per agent per round. At the start of each round, the backend resumes waiting groups whose members have all reached the same intermediate destination and may reset moving groups if they newly enter a coordinator influence zone. Only agents whose context has changed and are therefore in \texttt{DISCUSSING} are assembled into a deliberation workload. For each such agent, the system constructs a textual local view from cached geometry, nearby people density, active threats, and ranked exits; then it chooses either a group-discussion prompt or a solo-decision prompt depending on social role. Histories are trimmed to bounded windows, compact per-agent payloads are packed into batched requests, and asynchronous workers dispatch these requests to \texttt{gpt-4.1-nano} under configurable batch-size and concurrency limits. Returned destinations are normalized through fuzzy matching before they update the group destination tracker, and if a group exceeds the discussion budget the system forces a consensus destination using prior votes or nearest-exit fallback. This design is central to the system's real-time scale because it limits LLM work to moments where the social context actually changed.

\begin{whatifimplalgorithm}{Processing the Deliberation Stage of a Round}
\REQUIRE live simulation instance $\mathcal{S}$ at round $r$
\ENSURE updated decisions, messages, and targets for agents that re-deliberate
\STATE Resume waiting groups whose members all arrived at the same intermediate destination
\STATE Detect coordinator influence on moving groups and reset influenced agents to \texttt{DISCUSSING}
\STATE Collect the current set of agents in state \texttt{DISCUSSING}
\STATE Build one round-wide position array for fast nearby-people queries
\FORALL{discussing agents}
  \STATE Build a cached surroundings description from map geometry, nearby agents, threats, and exits
  \IF{agent acts alone}
    \STATE Prepare a solo decision prompt with local context and optional coordinator directive
  \ELSE
    \STATE Prepare a group discussion prompt with recent group chat, arrival status, and local context
  \ENDIF
  \STATE Trim history and pack a compact per-agent batch request
\ENDFOR
\STATE Partition requests into batches of size $B$ and dispatch async workers
\FORALL{responses returned by the workers}
  \STATE Normalize structured JSON output and fuzzy-match the destination string
  \STATE Update agent rationale, message history, and current target if a destination was chosen
  \IF{the response contains a destination}
    \STATE Record the validated destination in the group destination tracker
  \ENDIF
\ENDFOR
\STATE Force any overlong group discussion to a consensus destination if the round budget is exceeded
\end{whatifimplalgorithm}

\subsection{Unified Intervention Processing Across Modalities}
All steering operations described in the main paper flow through one shared intervention layer. Direct manipulation on the canvas, sidebar tool actions, typed requests in the What-If Agent, and spoken requests routed through the voice pipeline all end up as the same structured \texttt{(action, data)} commands. The natural-language path is implemented as a grounded translation step: ambient speech is first filtered for intervention intent, then \texttt{agent\_interpret} uses the current round, active environment, and existing overlays to translate vague requests such as ``place another coordinator near the south side'' into executable commands with concrete parameters. On the backend, \texttt{handle\_intervention} validates each command and dispatches it to a typed handler. Handlers mutate only the affected subsystem: announcements rebuild every active agent's prompt context, coordinator edits rebuild influence zones, exit and obstacle changes rebuild navigation and destination caches, and threat insertions update both the environment overlay and the threat-specific context injected into nearby agents. Because public interventions can invalidate prior plans, handlers also reset affected agents into discussion and clear stale group-tracker state, yielding one coherent control plane across all interaction modalities.

\begin{whatifimplalgorithm}{Applying an Intervention Request}
\REQUIRE input event $e$ from canvas manipulation, tool panel, typed text, or voice
\ENSURE updated shared environment state and agent context
\IF{$e$ is natural language}
  \STATE Filter ambient speech for intervention relevance when needed
  \STATE Translate the utterance into one or more structured commands using current simulation context
\ELSE
  \STATE Convert the UI gesture directly into one structured command
\ENDIF
\FORALL{commands $(action, data)$ extracted from $e$}
  \STATE Validate coordinates, names, counts, and threat types for $(action, data)$
  \STATE Dispatch to the typed intervention handler for $action$
  \IF{$action$ changes public context}
    \STATE Reset affected agents to \texttt{DISCUSSING} and rebuild prompt history when required
    \STATE Clear or rebuild group trackers, coordinator zones, navigation grids, and map caches as needed
  \ENDIF
  \STATE Persist the intervention record and refreshed runtime state
\ENDFOR
\STATE Broadcast updated positions, states, destinations, and environment overlays to subscribed collaborators
\end{whatifimplalgorithm}

\subsection{Deterministic Navigation and Hazard Dynamics}
The spatial engine executes the ``how'' of evacuation. Movement runs on a precomputed grid with flow fields toward exits and region destinations; each step blends global flow direction with direct-to-target motion, local neighbor repulsion, and threat repulsion, then resolves collisions by sliding along obstacles and probing alternative walkable directions. Speed is adjusted from crowd density, coordinator proximity, injury, and mobility settings, while a per-agent navigation state tracks stuckness and smoothed velocity to suppress oscillation in dense seating lanes. After movement, the engine marks arrivals, transitions exit reachers to \texttt{EXITED}, and returns non-exit arrivals to \texttt{WAITING} until their group is ready to deliberate again. Hazard dynamics are then advanced in place: fire radii expand, shooter threats follow a discrete behavioral state machine with movement and firing phases, police pursue the nearest active shooter, and both stampede and direct-threat fatalities are processed before the next round. This deterministic layer is what keeps WhatIf responsive while still surfacing interpretable crowd dynamics.

\begin{whatifimplalgorithm}{Advancing Movement and Hazards for One Round}
\REQUIRE live simulation instance $\mathcal{S}$ after deliberation
\ENSURE updated positions, arrivals, hazards, and fatalities
\STATE Publish the currently active threats to the navigation layer
\STATE Compute local crowd density for all moving agents
\STATE Refresh the position cache used for neighbor-repulsion queries
\FORALL{moving agents in parallel}
  \STATE Compute movement speed from density, coordinator relief, injury, and mobility settings
  \STATE Obtain a navigation vector by blending flow-field guidance, direct target motion, local repulsion, and threat repulsion
  \STATE Resolve collisions against the walkable grid and clamp the result to valid space
  \IF{arrival tolerance is reached}
    \IF{destination is an exit}
      \STATE Mark the agent \texttt{EXITED}
    \ELSE
      \STATE Mark the agent \texttt{WAITING} and record arrival in the group destination tracker
    \ENDIF
  \ENDIF
\ENDFOR
\STATE Process stampede effects for high-density hazard situations
\STATE Advance each dynamic threat one step (e.g., fire spread, shooter state transition, police pursuit)
\STATE Apply fatalities and injuries, then invalidate any cached local descriptions affected by hazard changes
\STATE Mark the simulation complete if no agents remain active
\end{whatifimplalgorithm}

\subsection{Collaborative Synchronization}
Collaboration is organized around project-scoped WebSocket rooms. The frontend maintains one room connection with heartbeat, reconnect, presence, cursor, and annotation channels, while the backend represents each participant as a \texttt{CollaborativeSession}. Read-only ephemeral traffic such as cursor motion and sketches is relayed immediately, but mutating events such as stepping, interventions, initialization, and forking acquire a room lock because they touch shared simulation state and module-level navigation globals. Before executing any mutating command, the session restores the selected simulation's snapped environment; after the command, it snapshots the new environment back into the instance and broadcasts the resulting state only to users subscribed to that run. This arrangement lets multiple users inspect and steer one authoritative simulation without drifting into client-specific copies, while still allowing several runs to coexist inside the same project workspace. In implementation terms, collaboration is therefore not layered on top of the simulator as a separate feature; it is built into the runtime protocol that mediates every state mutation.

\begin{whatifimplalgorithm}{Handling a Collaborative Runtime Command}
\REQUIRE incoming WebSocket message $m$ from collaborator $u$
\ENSURE consistent shared state for all users subscribed to the target run
\IF{$m$ is cursor or annotation traffic}
  \STATE Relay the event immediately without taking the simulation lock
  \STATE \RETURN
\ENDIF
\IF{$m$ requests pause}
  \STATE Cooperatively cancel the autoplay loop, clear the pause gate, and broadcast paused state
  \STATE \RETURN
\ENDIF
\STATE Acquire the room lock for the target simulation
\STATE Restore the simulation's snapped environment globals before mutation
\IF{$m$ requests initialization}
  \STATE Create or restore the target live simulation instance
\ELSIF{$m$ requests one step or an autoplay tick}
  \STATE Advance one simulation round and emit progress updates
\ELSIF{$m$ requests an intervention}
  \STATE Apply the unified intervention pipeline
\ELSIF{$m$ requests a fork}
  \STATE Create a child simulation from the selected prior round
\ENDIF
\STATE Snapshot the environment globals back into the live simulation instance
\STATE Release the room lock
\STATE Broadcast the resulting state to all collaborators subscribed to that run
\end{whatifimplalgorithm}

\subsection{Snapshot-Based Replay, Forking, and Recap Generation}
To support replay, comparison, and counterfactual branching, WhatIf persists both coarse snapshots and exact runtime state in SQLite. After initialization and every completed round, the backend saves per-agent positions, states, environment overlays, decisions, and group messages, as well as a richer runtime payload containing prompt histories, targets, exposure counters, tracker maps, and per-simulation environment attributes. Loading a prior run can therefore produce either a frozen recap state or a fully live instance that resumes from the exact saved round. Forking creates a new simulation record with a parent pointer, copies the parent snapshots and interventions up to the selected round, restores exact runtime state when it exists for that round, and then continues execution from that shared prefix. On the client side, the recap pipeline hydrates snapshots and interventions and derives exit usage, congestion hotspots, mobility traces, threat trajectories, and chat highlights into a structured report used for single-run inspection and side-by-side comparison. In other words, the same persistence model that supports live steering also underwrites the paper's claims about temporal navigation, branching, and multi-level interpretability.

\begin{whatifimplalgorithm}{Persisting History and Building Recap Artifacts}
\REQUIRE simulation identifier $s$, optional fork round $r$
\ENSURE persistent data for replay, branching, and recap views
\FORALL{completed rounds of simulation $s$}
  \STATE Save a snapshot containing positions, states, state summary, environment, decisions, and group messages
  \STATE Save an exact runtime payload containing histories, targets, exposures, tracker maps, and simulation attributes
\ENDFOR
\IF{a user loads a prior run}
  \STATE Rehydrate either a frozen recap state or a live runnable instance from the saved snapshot and runtime payload
\ENDIF
\IF{a user forks at round $r$}
  \STATE Create a new simulation record with parent metadata and fork round $r$
  \STATE Copy parent snapshots and interventions up to $r$
  \STATE Restore runtime at $r$ when available and continue execution from that state
\ENDIF
\STATE Hydrate snapshot and intervention history in the frontend recap pipeline
\STATE Derive recap metrics: exit usage, congestion grids, mobility traces, threat trajectories, and chat highlights
\STATE Serialize the derived report for single-run recap and cross-run comparison
\end{whatifimplalgorithm}

\end{document}